\documentclass[pra,twocolumn]{revtex4-1}

\usepackage{graphicx}
\usepackage{epstopdf} 
\DeclareGraphicsExtensions{.eps, .pdf}
\usepackage{amssymb, amsmath}
\usepackage{hyperref}

\begin{document}

\newcommand{\bra}[1]{\ensuremath{\langle#1 |}}
\newcommand{\ket}[1]{\ensuremath{ |#1\rangle}}
\newcommand{\bracket}[2]{\ensuremath{\langle#1 \vphantom{#2} |  #2 \vphantom{#1} \rangle}}

\def\O{{\cal{O}}}
\def\Y{{\cal{Y}}}
\def\t{\theta}
\def\H{\tilde{H}}
\def\E{\tilde{E}}
\def\P2{\tilde{P}^2}
\def\X2{\tilde{X}^2}
\def\tphi{\tilde{\phi}}
\def\sinc{\text{sinc}}
\def\rect{\text{rect}}
\def\dH{\tilde{\cal{H}}}
\def\hphi{\hat{\phi}}
\def\bP{\mathbf{P}}
\def\bp{\mathbf{p}}
\def\bR{\mathbf{R}}
\def\br{\mathbf{r}}

\title{Digital quantum computation of fermion-boson interacting systems}
\author{Alexandru Macridin, Panagiotis Spentzouris, James Amundson, Roni Harnik }
\affiliation{Fermilab, P.O. Box 500, Batavia, Illinois 60510, USA}

\begin{abstract}

We introduce a new method for representing
the low energy subspace of a bosonic field theory on the qubit space of digital quantum computers. 
This discretization leads to an exponentially precise description of the subspace of the continuous theory thanks to the Nyquist-Shannon sampling theorem.
The  method makes the implementation of quantum algorithms
for purely bosonic systems as well as fermion-boson interacting systems feasible. 
We present  algorithmic circuits  for  computing the time evolution of these systems.  
The  complexity of the algorithms scales polynomially with the system size.
The  algorithm 
is a natural extension of the  existing quantum algorithms for 
simulating fermion systems in quantum chemistry and condensed matter physics to
systems involving bosons and fermion-boson interactions and has a broad variety of
potential applications in particle physics, condensed matter, etc.
Due to the relatively small amount of additional  resources 
required by the inclusion of bosons in our algorithm, the simulation of electron-phonon and similar systems
can be placed  in the same near-future reach as the simulation of interacting electron systems.
We benchmark our algorithm by implementing it for a $2$-site Holstein polaron problem on an Atos Quantum Learning Machine (QLM) quantum simulator. 
The polaron quantum simulations are in excellent agreement with the results obtained by exact 
diagonalization.

\end{abstract}

\maketitle

\section{Introduction}
Recent advances in quantum hardware technology have initiated a new era in computing science.
As quantum hardware has advanced, the development 
of quantum algorithms has become an area of intensive research.
Quantum computers are naturally suited to simulate the evolution of quantum systems.
For example,  the algorithms for simulating  fermion systems
in quantum chemistry and condensed matter physics have proven to be especially 
successful~\cite{abrahams_lloyd_prl_1997,abrahams_lloyd_prl_1999, ortiz_pra_2001,somma_gubernatis_2002,somma_qic_2003,
whitfield_2011,troyer_pra_2015,peruzzo_nature_2014,mcclean_NJPhys_2016}.
Due to the relatively small amount  of  resources required, optimized fermion algorithms
are very promising for near-future quantum simulations. Unfortunately, purely fermionic
models preclude the simulation of physically important theories with bosonic degrees
of freedom such as phonons,
photons and gluons, which appear in condensed matter and high energy physics.
In this paper we extend the existing fermion algorithms to include bosons, opening up the possibility for quantum simulation
to whole new classes of physical systems.

This paper addresses non-relativistic fermion-boson quantum field theories with focus
on electron-phonon systems. The interaction of electrons  with other bosonic collective
excitations in solids (such as spin, orbital, charge, etc.)\ can be addressed  
by models similar to the electron-phonon model. 
We address both fermion-boson and  boson-boson interactions.
Our algorithm can also be applied  to quantum optics problems. 
Since the quantum simulation of relativistic field theories is as an important
goal for high energy physics, we consider this approach as a first step  
towards that direction.

While there are established ways to map fermion states to qubits~\cite{ortiz_pra_2001, whitfield_2011,kitaev_bravy},
the literature contains fewer discussions on representing  bosons on gate quantum computers.
As discussed in Ref~\cite{Lidar}, bosons can be 
represented as a sum  of $n_x$  parafermions (qubits), 
up to an error  $\O(n/n_x)$, where $n$ is the boson state occupation number.  This representation requires a large number 
of qubits, especially in the intermediate and strong coupling regimes where $n$ is large. In addition, no algorithm
has been proposed to describe the evolution of boson states in this representation. 
In Refs.~\cite{somma_qic_2003,batista_ortiz_2007} purely bosonic systems with a fixed number of bosons are addressed,
but the method is not suitable for fermion-boson interacting systems where the number of bosons is not conserved.
An algorithm for calculating scattering amplitudes  in  quantum field theories has been  proposed in Ref~\cite{jordan_science_2012}. 
Their approach  is based on the discretization of the continuous field value at each lattice site.  
The required number of qubits per lattice site scales as  $\log(1/\epsilon)$, where $\epsilon$
is the desired accuracy. Our algorithm also relies on field discretization, but the number of qubits per lattice site needed to
represent the bosons
scales exponentially faster, $\approx \log(\log(1/\epsilon))$.  In fact, 
when our algorithm is applied to electron-phonon models, we find that
only a small number of 
additional qubits per site,  $n_x \approx 6\rm{~or~}7$, is enough  to simulate phonons with exponentially good accuracy in most 
problems of physical interest, including the weak, intermediate, and strong coupling regimes. 

It is worth mentioning that the quantum computation of interacting fermion-boson systems 
has been addressed  
in trapped ion systems~\cite{trapped_ions1,trapped_ions2,trapped_ions3,trapped_ions4}. 
In these cases the bosonic states were 
mapped to the ions' vibrational states. However, this method
is specific to the particular kind of hardware used, which
possesses additional degrees of freedom ({\em i.e.,}  vibrational states of ions) in
addition to qubits; the additional degrees of freedom
were used for the boson representation. Our approach to quantum computation of
systems with bosons is different, since  we consider
boson representation solely on qubits.

The representation of the boson space on qubits is the most important result of this work.
This representation allows an efficient simulation of the evolution operator of the
fermion-boson systems.
The bosonic degrees of freedom are treated as a finite set of harmonic oscillators. 
We show that the low-energy space of a harmonic oscillator
is, up to an exponentially small error,  
isomorphic  with the low-energy subspace of a finite-sized Hilbert space. 
The finite-sized boson Hilbert space is mapped onto the qubit space  of universal quantum computers.
The size of the low-energy subspace is given by the maximum boson number cutoff; 
the finite size of the Hilbert space increases  linearly with this cutoff.
The number of qubits necessary to store the bosons scales logarithmically with the cutoff.

Our method for truncating the harmonic oscillator space can also be applied to
simulate  the Schrodinger equation on a quantum computer.
A similar finite-sized Hilbert space truncation is  employed by the Fourier grid Hamiltonian 
(FGH) method~\cite{ marston_kurti_jcphys_1989} and is related to more general discrete variable representation  
(DVR) methods~\cite{light_jcphys_1985, Littlejohn_2002,bulgac_forbes_prc_2013}. 
We present a novel explanation for the exponential accuracy of the FGH method 
based on the Nyquist-Shannon sampling theorem~\cite{nyquist-shanon}.

The fermions in our algorithm are mapped to qubit states via the Jordan-Wigner transformation~\cite{jordan_wigner_1928,ortiz_pra_2001, whitfield_2011}.
Quantum algorithms for interacting fermions have been addressed at length in numerous papers; 
see for example Refs.~\cite{somma_gubernatis_2002, whitfield_2011,troyer_pra_2015}.
The evolution of the pure-fermion Hamiltonian is not addressed here.
We present algorithmic circuits for  the evolution of  the pure-boson Hamiltonian 
and the fermion-boson interacting Hamiltonian. 
The additional qubits needed to accommodate bosons is $\O(N n_x)$  where $N$ is the number of harmonic oscillators,
which scales linearly with the system size,
and $n_x$ is the number of qubits per harmonic oscillator, which is independent of $N$.
For long-range $m$-body boson-boson interactions ({\em i.e.,} an $m$-leg interaction vertex),
the additional circuit depth is $\O(N^m)$.
In general long-range fermion-boson interactions yield an additional depth of $\O(N^2)$. 
However, when the bosons couple to the fermion hopping, as happens in electron-phonon models,
the  additional depth scales as  $\O(N)$.
For finite range boson-boson and fermion-boson interaction the additional circuit depth is constant.

As an example of fermion-boson interacting systems  we address the polaron problem~\cite{landau_polaron}. The polaron is a bound state 
between an electron and its induced crystal lattice deformation; it can be thought of as an electron 
dressed by phonons. Although it involves only one electron, the polaron problem is nontrivial
and in general cannot be solved on classical computers due to the exponential increase of the
Hilbert space with increasing system size. Polaronic effects can significantly change 
the electric and transport properties of materials,  angle resolved-photoemission spectra, superconducting properties, etc.
We benchmark our algorithm by running  a simulation of the two-site  Holstein  polaron~\cite{holstein_1959},
utilizing the Quantum Phase Estimation (QPE) method~\cite{kitaev_qpe,cleve_1998, abrahams_lloyd_prl_1999, kitaev_qpe_2002, nielsen_2010, guzik_science_2005} 
on an Atos Quantum Learning Machine simulator. The energy and phonon distribution of the polaron state
agree with results obtained from exact diagonalization.

The paper is organized as follows. In Section~\ref{sec:fbham} the
fermion-boson model is introduced.
 In Section~\ref{sec:bosons} we address the representation of bosons on a finite-sized space. 
The quantum algorithm
is described in Section~\ref{sec:algor}.
Section~\ref{sec:polaron} presents the results of the QPE simulation for
the Holstein polaron. In Section~\ref{sec:discuss}
we discuss the general applicability of our approach to physical systems.
Summary and conclusions are given  in Section~\ref{sec:concl}.

\section{Fermion-boson Hamiltonian}
\label{sec:fbham}

In our algorithm  the fermion operators
appearing in the Hamiltonian need to be expressed in the second quantized form.  On the other hand, the bosonic 
operators are required to be written as function of the canonical ``position'' and ``momentum''
operators $X$ and $P$, obeying the  commutation relation  $[X,P]=i$. 

In this section we  start with  the electron-phonon Hamiltonian, since it constitutes one of the most common 
physical examples of non-relativistic fermion-boson interacting systems. We will follow 
with a general fermion-boson Hamiltonian written in the second quantized form and will
describe the steps necessary to rewrite it in a form suitable for quantum computation.

\subsection{Electron-phonon model}
\label{sec:ep_model}

The electron-phonon model describes the electronic and ionic degrees of freedom 
in a solid. The model can be derived (see Refs~\cite{alexandrov_elph, kitel_qts, giustino_rmp_2017} for more details)
from the Hamiltonian

\begin{eqnarray}
\label{eq:gen_elphham}
H&=&\sum_{i} \frac{p^2_i}{2 m}  +\sum_{n,\alpha} \frac{P^2_{n \alpha}}{2 M_\alpha} + \sum_{i \ne j} V_e\left(r_i,r_j\right)\\ \nonumber 
 &+& \sum_{n \alpha \ne m \beta} V_p \left(R_{n \alpha},R_{m \beta}\right)  + \sum_{i,n \alpha} V_{ep} \left(r_i,R_{n \alpha}\right).
\end{eqnarray}
\noindent In Eq.~(\ref{eq:gen_elphham}) the  index $i$  labels the electrons while the index $n$  labels the crystal's unit cells.
The ions in the unit cell are labeled by $\alpha$.
The first two terms  represent the kinetic energy of the electrons and  ions, while the
last three terms describe the electron-electron, ion-ion and electron-ion interactions, respectively.

With the assumption that the ions' motion is characterized only by small 
displacements around their equilibrium position $R_0=\{R_{n \alpha 0}\}_{n \alpha}$,
Eq.~(\ref{eq:gen_elphham}) can be written as
\begin{equation}
\label{eq:gen_elham}
H=H_e+H_{p}+H_{ep},
\end{equation}
\noindent with
\begin{eqnarray}
H_e&=&\sum_{i} \frac{p^2_i}{2 m}  + \sum_{i} V_{ep} \left(r_i,R_0 \right) + \sum_{i \ne j} V_e\left(r_i,r_j\right),  \\  
\label{eq:hpv}
H_{p}&=& \sum_{n \alpha} \frac{P^2_{n \alpha}}{2 M_\alpha}
+\sum_{n \alpha, m \beta}\frac{\partial^2 V_p\left( R_0 \right)}{\partial R_{n \alpha} \partial R_{m \beta}}\Delta R_{n \alpha} \Delta R_{m \beta},\\ 
H_{ep}&=& \sum_{i,n \alpha} \frac{\partial V_{ep}\left(r_i, R_0 \right)}{\partial R_{n \alpha}}  \Delta R_{n \alpha}.
\end{eqnarray}
\noindent Since the ions' potential energy is minimum at the equilibrium position $R_0$, we have taken $\partial V_p(R_0)/\partial R_{n \alpha} =0$
when deriving Eq.~(\ref{eq:hpv}).

The term $H_e$ contains only the electronic degrees of freedom and reads in the second quantized form
\begin{equation}
\label{eq:hame}
H_e= \sum_{ij} t_{ij} \left( c^{\dagger}_i c_j +c^{\dagger}_j c_i  \right) + \sum_{ijkl} U_{ijkl}  c^{\dagger}_i c^{\dagger}_j c_k c_l,
\end{equation}
\noindent where  $c^{\dagger}_i$ ($c_i$) represents the electron creation (annihilation) operator for the state $i$. 

The term $H_p$ describes the ionic vibration and can be written as a sum of coupled harmonic oscillators,
\begin{equation}
\label{eq:hamp}
H_p = \sum_{n\nu} \frac{P^2_{n\nu}}{2 M_{\nu} } + \frac{1}{2} M_{\nu}\omega^2_{n\nu} X^2_{n\nu} + \sum_{n\nu m \mu}K_{n\nu m\mu} X_{n\nu} X_{m\mu},
\end{equation}
\noindent where $\nu$ and $\mu$ are vibrational mode labels. 
The operators $X_{n\nu} = {\bf{O}} (\{\Delta R_{n \alpha}\})$
and $P_{n\nu} = {\bf{O}} (\{P_{n \alpha }\})$ obey the canonical commutation relation
$[X_{n\nu},P_{m\mu}]=i \delta_{nm}\delta_{\nu \mu}$ and 
are obtained by  an orthogonal transformation ${\bf{O}}$ of the  vectors
$\{\Delta R_{n \alpha }\}$ and  $\{P_{n \alpha}\}$, respectively.
In general the vibrational modes  are  determined by requesting
that  the Hamiltonian~(\ref{eq:hamp}) written in the momentum basis reduces to a sum of independent
oscillators. Since coupled harmonic oscillators can be easily simulated  with our algorithm,
for our purpose this decomposition into independent momentum modes 
is not necessary and in most cases not even optimal.
The mode label $\nu$ in Eq.~(\ref{eq:hamp}) represents  
just a convenient basis choice. The optimal basis is dependent on the particular system under
investigation. As will become clear later, the algorithm
is efficient in a basis where the interactions have short range and the number of phonons per state is small.

The electron-ion interaction term is
\begin{equation}
\label{eq:hamep}
H_{ep} = \sum_{ij n \nu} g_{ijn\nu} \left( c^{\dagger}_i c_j +c^{\dagger}_j c_i  \right) X_{n \nu},
\end{equation}
\noindent and couples  single-particle electron operators with ions' position operators.

Note that in the literature, unlike in our representation, both the electron and  
phonon operators in the electron-phonon Hamiltonians are usually written 
in the second quantized form.

\subsection{General fermion-boson model}
\label{sec:sfb_model}

We start with a fermion-boson Hamiltonian written in the second quantized form,
\begin{eqnarray}
\label{eq:gen_model}
H=H_f+H_b+H_{fb}
\end{eqnarray}
\noindent where $H_f$ is the fermion Hamiltonian, as in Eq.~(\ref{eq:hame}),
$H_b$ contains only bosonic degrees of freedom and $H_{fb}$ describes the fermion-boson interaction.

\subsubsection{Boson Hamiltonian}
\label{ssec:bh}

We split the  boson Hamiltonian into three parts
\begin{eqnarray}
\label{eq:gen_bham}
H_b =H_{b0} +H_{bs}+H_{bi}.
\end{eqnarray}
The term $H_{b0}$ is noninteracting and is written
\begin{eqnarray}
\label{eq:b0ham}
H_{b0} =\sum_{mn} \xi_{m n} b^{\dagger}_m b_n+\sum_n \left( \zeta_n b^{\dagger}_n + \zeta_n^* b_n \right).
\end{eqnarray}
The term $H_{bs}$ is the {\em squeezing} Hamiltonian~\cite{garry_knight_book}; we consider it explicitly since
it is of interest in quantum optics.  It reads
\begin{eqnarray}
\label{eq:bsham}
H_{bs} =\sum_{nm} \left( \lambda_{nm} {b^{\dagger}_n} {b^{\dagger}_m} + \lambda_{nm}^* {b_n} {b_m} \right).
\end{eqnarray}
\noindent The Hamiltonian $H_{bi}$ contains interacting terms  consisting  of a product of three or four 
creation (or annihilation) operators,  
\begin{eqnarray}
\label{eq:b1ham}
H_{bi} &=&\sum_{nmr} U_{nmr} b^{\dagger}_n b^{\dagger}_m b_r +\sum_{nmr}  V_{nmr} b^{\dagger}_n b^{\dagger}_m  b^{\dagger}_r \\ \nonumber 
&+&\sum_{nmrs} T_{nmrs} b^{\dagger}_n b^{\dagger}_m b_r b_s + 
\sum_{nmrs} W_{nmrs} b^{\dagger}_n b^{\dagger}_m b^{\dagger}_r b_s\\ \nonumber 
&+&\sum_{nmrs}Y_{nmrs}  b^{\dagger}_n b^{\dagger}_m b^{\dagger}_rb^{\dagger}_s + h.c.
\end{eqnarray}
\noindent The notation $h.c.$\ means Hermitian conjugate and ensures the Hamiltonian (\ref{eq:b1ham}) is Hermitian.
In Eq.~(\ref{eq:b1ham}) we consider the  order of the interaction to be at maximum $4$, ({\em i.e.}, allowing for up to $4$-leg interaction vertices), as this is relevant for modeling gluon-gluon interactions in quantum chromodynamics.
However, higher-order interaction terms
can be considered  in our algorithm as well.

The boson creation and annihilation operators obey the commutation relations $[b_n, b^{\dagger}_m]=\delta_{nm}$,
$[b_n, b_m]=0$ and $[b^{\dagger}_n, b^{\dagger}_m]=0$. The following transformation
\begin{eqnarray}
\label{eq:b_to_x}
X_n&=& \frac{1}{\sqrt{2 l_n}}\left(b^{\dagger}_n+b_n\right) \\
\label{eq:b_to_p}
P_n&=& i\sqrt{\frac{l_n}{2}}\left(b^{\dagger}_n-b_n\right),
\end{eqnarray}
\noindent where $l_n$ is an arbitrary constant, yields canonical position and momentum  operators
(which are proportional to quadrature operators in quantum optics), satisfying $[X_{n},P_{m}]=i \delta_{nm}$.

The Hamiltonian $H_{b0}$ becomes 
\begin{eqnarray}
\label{eq:xp_b0ham}
H_{b0} &=&\sum_{n}  \frac{\xi_{nn}}{l_n} \left(\frac{P^2_n}{2}  + \frac{l^2_n}{2} X^2_n  -\frac{l_n}{2}\right)\\ \nonumber
&+&\sum_{m < n}  \Re \xi_{mn} \left(\frac{P_m P_n}{\sqrt{l_m l_n}}  + \sqrt{l_m l_n} X_m X_n \right)  \\ \nonumber
&+& \sum_{m < n}  \Im \xi_{mn} \left( -\sqrt{\frac{l_n}{l_m}}  X_n P_m+ \sqrt{\frac{l_m}{l_n}}  X_m P_n  \right)\\ \nonumber
&+& \sum_{n}  \Re \zeta_n \sqrt{2 l_n} X_n +\sum_{n}  \Im \zeta_n \sqrt{\frac{2}{l_n}} P_n.
\end{eqnarray}
\noindent The Hamiltonian (\ref{eq:xp_b0ham}), which is analogous to the phonon Hamiltonian (\ref{eq:hamp}),
consists of a sum of coupled harmonic oscillators. However, unlike Eq.~(\ref{eq:hamp}) where only coupling 
of the type $X_nX_m$ between the position operators at different sites is considered, 
the Hamiltonian (\ref{eq:xp_b0ham}) also includes coupling terms of type $P_nP_m$ and $X_nP_m$ ($n \ne m$).
Besides, the Hamiltonian (\ref{eq:xp_b0ham})  includes linear coupling of the canonical position and momentum
operators to some arbitrary fields $\Re \zeta$ and $\Im \zeta$, respectively. 

The squeezing Hamiltonian can be written as
\begin{eqnarray}
\label{eq:xp_bsham}
H_{bs} &=&\sum_{n \ne m} -\Re  \lambda_{nm} \left(\frac{P_n P_m}{\sqrt{l_n l_m}} - \sqrt{l_n l_m} X_n X_m \right)\\ \nonumber
&+&\sum_{n \ne m} \Im \lambda_{nm} \left(\sqrt{\frac{l_n}{l_m}} X_n P_m + \sqrt{\frac{l_m}{l_n}} P_n X_m \right)  \\\nonumber
&+&\sum_n - \frac{\Re \lambda_{nn}}{l_n}  \left(\frac{P^2_n}{2}  -\frac{l^2_n}{2} X^2_n \right)\\ \nonumber
&+&\frac{1}{2}\sum_n \Im \lambda_{nn}  \left( X_n P_n+P_n X_n \right).
\end{eqnarray}
\noindent As in $H_{b0}$~(\ref{eq:xp_b0ham}), the Hamiltonian~(\ref{eq:xp_bsham}) contains terms 
$X_n P_m$  ($n \ne m$), $X_n X_m$ and $P_n P_m$.  It also contains local
terms $X_n P_n$ which require a different algorithmic implementation, 
as explained in Section~\ref{ssec:ev_bosonboson}.

Employing Eqs.~(\ref{eq:b_to_x}) and (\ref{eq:b_to_p}), the Hamiltonian $H_{bi}$ (\ref{eq:b1ham}) transforms
into a sum of terms of type  $A_{n}A_{m}A_{r}$ and $A_{n} A_{m} A_{r} A_{s}$, where $A_{n}$ is either the
$X_{n}$ or the $P_{n}$ operator of the harmonic oscillator $n$.

\subsubsection{Fermion-boson coupling}

We consider a model where the interaction is given by coupling single-particle fermion operators 
with boson creation and annihilation operators,
\begin{eqnarray}
\label{eq:sq_fbham}
H_{fb} =\sum_{ijn} \left( g_{ijn} c^{\dagger}_i c_j b^{\dagger}_n +g^*_{ijn} c^{\dagger}_j c_i b_n \right).
\end{eqnarray}

After employing Eqs.~(\ref{eq:b_to_x}) and (\ref{eq:b_to_p}), $H_{fb}$ becomes
\begin{eqnarray}
\label{eq:fbham1}
H_{fb} &=&\sum_{ijn} \sqrt{\frac{l_n}{2}} \Re g_{ijn} \left( c^{\dagger}_i c_j +  c^{\dagger}_j c_i \right) X_n\\ 
\label{eq:fbham2}
&+& \sum_{ijn} \frac{-\Im g_{ijn}}{\sqrt{2 l_n}}  \left(c^{\dagger}_i c_j + c^{\dagger}_j c_i \right) P_n\\ 
\label{eq:fbham3}
&+& \sum_{ijn} i \sqrt{\frac{l_n}{2}} \Im g_{ijn} \left( c^{\dagger}_i c_j - c^{\dagger}_j c_i \right) X_n\\ 
\label{eq:fbham4}
&+& \sum_{ijn} \frac{i\Re g_{ijn}}{\sqrt{2 l_n}} \left(c^{\dagger}_i c_j -  c^{\dagger}_j c_i \right) P_n.
\end{eqnarray}
\noindent The first term above, Eq.~(\ref{eq:fbham1}),  is of the same type as the electron-phonon 
coupling,  Eq.~(\ref{eq:hamep}). The second term, Eq.~(\ref{eq:fbham2}), represents the coupling
of the fermion kinetic energy operator to the boson momentum operator.
The last two terms, Eq.~(\ref{eq:fbham3})  and Eq.~(\ref{eq:fbham4}), describe
the coupling of the fermion current operator to the boson 
position and  momentum  operators, respectively.

\section{Boson representation on a finite space}
\label{sec:bosons}

The boson operators in our model are the  canonical position and momentum operators, 
as discussed in Section~\ref{sec:fbham}.  The bosons are described by a set of 
coupled harmonic oscillators labeled by state index (for example, the position and vibrational mode indices for phonons).
The boson Hilbert space is a direct product of the Hilbert spaces of the harmonic oscillators. 
In this section we  address the representation of the harmonic oscillator space on a finite-sized space.
In Section~\ref{sec:algor} we will show how to map this finite-sized space onto the qubit space of a quantum computer.

\subsection{Harmonic oscillator}
\label{ssec:ho}

The harmonic oscillator is described by the Hamiltonian
\begin{eqnarray}
\label{eq:hos_ham}
H_h=\frac{1}{2}P^2+\frac{1}{2}X^2,
\end{eqnarray}
\noindent where the operators $X$, $P$ and $H_h$ are rescaled by $1/\sqrt{M \omega}$, $\sqrt{M \omega}$ and
 $1/\omega$, respectively. The   eigenvalues  and eigenvectors of $H_h$ are
\begin{equation}
\label{eq:hos_en}
E_n = n+\frac{1}{2}
\end{equation}
and
\begin{equation}
\label{eq:phi_n}
\ket{\phi_n}  =  \int \phi_n(x) \ket{x} dx=  \int \hat{\phi}_n(p) \ket{p} dp, 
\end{equation}
\noindent  where  $\{\ket{x}\}$ is the position coordinate basis and $\{\ket{p}\}$ is the
momentum coordinate basis. The eigenfunction 
\begin{eqnarray}
\label{eq:HGx}
\phi_n(x) = \frac{1}{\pi^{\frac{1}{4}}\sqrt{2^n n!}} e^{-\frac{x^2}{2}}H_n(x)
\end{eqnarray}
\noindent is the  Hermite-Gauss (HG) function of order $n$ and $\hat{\phi}_n(p)$ is its Fourier transform
to the momentum representation.

In addition to being eigenfunctions of the harmonic oscillator Hamiltonian, the HG functions are 
also eigenfunctions of the Fourier transform operator~\cite{FT_HG}, 
\begin{eqnarray}
\label{eq:fthg}
[{\cal{F}} (\phi_n)](p) \equiv \hat{\phi}_n(p)=(-i)^n\phi_n(p).
\end{eqnarray}

The HG functions  satisfy the relations
\begin{equation}
\label{eq:xhg}
x\phi_n(x)=\frac{1}{\sqrt{2}}\left( \sqrt{n+1} \phi_{n+1}(x) + \sqrt{n} \phi_{n-1}(x) \right)
\end{equation}
and
\begin{equation}
\label{eq:phg}
p \hat{\phi}_n(p)=\frac{i}{\sqrt{2}}\left( \sqrt{n+1}   \hat{\phi}_{n+1}(p) - \sqrt{n} \hat{\phi}_{n-1}(p) \right).
\end{equation}
\noindent Eq.~(\ref{eq:xhg}) follows from the recurrence relations of the Hermite polynomials~\cite{recurence_hn}, while
Eq.~(\ref{eq:phg}) can be obtained from Eq.~(\ref{eq:xhg}) by employing 
Eq.~(\ref{eq:fthg}). Note that Eqs.~(\ref{eq:xhg}) and (\ref{eq:phg}) are just the familiar eigenvalue equations 
$\bra{x}X\ket{\phi_n}=x \bracket{x}{\phi_n}$ and  $\bra{p}P\ket{\phi_n}=p \bracket{p}{\phi_n}$, respectively, where the position 
operator $X= \left( b^{\dagger}+ b \right)/\sqrt{2}$
and momentum operator $P= i\left( b^{\dagger} - b \right)/\sqrt{2}$ are written as functions of the creation and the 
annihilation operators $b^{\dagger}$ and $b$. Remember that the operators $b^{\dagger}$ and $b$ satisfy
\begin{eqnarray}
\label{eq:bphg}
b^{\dagger}\ket{\phi_n}&=&\sqrt{n+1}\ket{\phi_{n+1}}, \\
\label{eq:bhg}
 b\ket{\phi_n}&=&\sqrt{n}\ket{\phi_{n-1}}.
\end{eqnarray}

The equations (\ref{eq:xhg}) and (\ref{eq:phg})  together with the exponential decay at large argument, 
Eqs.~(\ref{eq:HGx}) and (\ref{eq:fthg}), are the essential properties of the HG functions which make the 
controlled truncation of the Hilbert space discussed in Section~\ref{ssec:dho} feasible.

\subsection{Discretization of the harmonic oscillator space}
\label{ssec:dho}

In both the coordinate and momentum representations the HG functions decay exponentially quickly to zero for large arguments.
The width of the HG functions, {\em i.e.,} the interval range  enclosing most of the weight $|\phi_n(x)|$,  
increases  with increasing $n$.
For a cutoff number $N_{ph}$ one can define a width $2L$ such that 
$|\phi_n(x)| \approx 0$ when $|x|>L$ for all $n<N_{ph}$. Eq.~(\ref{eq:fthg}) implies that for the same
$L$, $|\hat{\phi}_n(p)| \approx 0$ when $|p|>L$ and $n<N_{ph}$.
The error of these approximations can be made arbitrarily small through a large enough choice of $L$.

Restricting the problem to the region $|p|<L$, the Nyquist-Shannon sampling theorem~\cite{nyquist-shanon} 
for band limited signals applies. It shows that, without  loss of information,  $\phi_n(x)$ can be sampled at 
points $x_i=i \Delta$, where $i$ is an integer
and $\Delta=\pi/L$. Moreover, with the same exponentially small error, we can restrict $i$ 
to $N_x$ points such that  $x_i \in [-L,L]$.
The HG function can be written as
\begin{eqnarray}
\label{eq:nysh}
\phi_n(x)=\sum_{i=-\frac{N_x}{2}}^{\frac{N_x}{2}-1} \phi_n(x_i) u_i(x)+ \O(\epsilon),
\end{eqnarray}
\noindent where $u_i(x)=\sinc\left( \frac{x-x_i}{\Delta} \right)$.  The minimum $N_x$ to satisfy the requirement
$x_i \in [-L,L]$ is given by the equation $2L = N_x \Delta$, which implies
$2L = \sqrt{2 \pi N_x}$ and $\Delta=\sqrt{2 \pi /N_x}$ with $i=\overline{-N_x/2, N_x/2-1}$.
For a more detailed discussion about 
the application of the Nyquist-Shannon sampling theorem to the HG functions  see  Appendix~\ref{app:hor}.

The properties of the HG functions  require that $N_x>N_{ph}$. This can
be understood within the framework of the WKB approximation~\cite{merzbacher} for the
harmonic oscillator. The 
magnitude of $\phi_n(x)$ is exponentially small in the classically forbidden region
where $V(x)-E_n>0$, and enhanced around the turning points defined by $V(x_{tn})-E_n=0$.
In the harmonic oscillator potential $V(x)=x^2/2$ 
the turning points are given by $x_{tn}=\pm \sqrt{ 2 n +1}$.
The condition $L>|x_{tn}|$ for $n =N_{ph}$ required by the Nyquist-Shannon sampling theorem implies
\begin{eqnarray}
\label{eq:NxNph}
 N_x   >  \left(\frac{4}{\pi} N_{ph} +\frac{2}{\pi} \right),
\end{eqnarray}
\noindent {\em i.e.,} $N_x>N_{ph}$.

Let us consider the finite-sized subspace $\tilde{\cal{H}}$ spanned by the sampling  position vectors 
$\{ \ket{x_i} \}_i$ for integer $i=\overline{-N_x/2, N_x/2-1}$,  and 
define the vectors $\ket{\chi_n} \in \dH$ by
\begin{eqnarray}
\label{eq:dhg}
\bracket{x_i}{\chi_n} \equiv \sqrt{\Delta} \phi_n(x_i).
\end{eqnarray}
Two essential properties of the vectors $\ket{\chi_n}$ with $n<N_{ph}$ are an immediate consequence of the
Nyquist-Shannon sampling theorem. First, 
the vectors $\ket{\chi_n}$ 
are orthonormal; see Eqs.~(\ref{eq:dhg_orto}) and (\ref{eq:adhg}).
Second (see Eqs.~(\ref{eq:adft}) and (\ref{eq:adft1})) 
\begin{eqnarray}
\label{eq:pdhg}
\bracket{p_m}{\chi_n}=\sqrt{2 \pi \Delta}\hat{\phi}_n(p_m),
\end{eqnarray}
\noindent where the vectors
\begin{eqnarray}
\label{eq:oFFT_xp}
 \ket{p_m} = \frac{1}{\sqrt{N_x}}\sum_{i=-\frac{N_x}{2}}^{ \frac{N_x}{2}-1 } e^{i x_i p_m} \ket{x_i},
\end{eqnarray}
\noindent with  $p_m=m \Delta$ and $m=\overline{-N_x/2, N_x/2-1}$, 
are obtained by  the discrete Fourier transform of the 
$\{\ket{x_i}\}$ vectors.
Equation (\ref{eq:dhg}) implies that $\bracket{x_i}{\chi_n}$ is proportional to the HG function $\phi_n(x)$
at the grid points $\{x_i\}$,
while Eq.~(\ref{eq:pdhg}) implies that its discrete Fourier transform  $\bracket{p_m}{\chi_n}$
is proportional to the HG functions in the momentum representation, $\hat{\phi}_n(p)$, at the grid points  $\{p_m\}$.
Therefore,  employing Eqs. (\ref{eq:xhg}) and (\ref{eq:phg}), one gets
\begin{equation}
\label{eq:xchi}
x_i \bracket{x_i}{\chi_n}=\frac{1}{\sqrt{2}}\left( \sqrt{n+1} \bracket{x_i}{\chi_{n+1}} + \sqrt{n} \bracket{x_i}{\chi_{n-1}} \right),
\end{equation}
\begin{equation}
\label{eq:pchi}
p_m \bracket{p_m}{\chi_n}=
\frac{i}{\sqrt{2}}\left( \sqrt{n+1} \bracket{p_m}{\chi_{n+1}} - \sqrt{n} \bracket{p_m}{\chi_{n-1}} \right).
\end{equation}

Now we define the discrete position and momentum operators acting on $\dH$ by
\begin{eqnarray}
\label{eq:tilde_x}
\tilde{X} \ket{x_i} =x_i \ket{x_i}, \\
\label{eq:tilde_p}
\tilde{P} \ket{p_m} =p_m \ket{p_m}.
\end{eqnarray}
\noindent  The equations (\ref{eq:xchi}) and (\ref{eq:pchi}) can be written as
\begin{eqnarray}
\label{eq:txchi}
\tilde{X} \ket{\chi_n}=\frac{1}{\sqrt{2}}\left( \sqrt{n+1} \ket{\chi_{n+1}} + \sqrt{n} \ket{\chi_{n-1}} \right),\\
\label{eq:tpchi}
\tilde{P} \ket{\chi_n}=\frac{i}{\sqrt{2}}\left( \sqrt{n+1} \ket{\chi_{n+1}} - \sqrt{n} \ket{\chi_{n-1}} \right),
\end{eqnarray}
\noindent which implies that 
\begin{eqnarray}
\label{eq:dxpcomphi}
[\tilde{X}, \tilde{P}] \ket{\chi_n}=i \ket{\chi_n} ~~\text{for}~~n<N_{ph}.
\end{eqnarray}
If we restrict to the subspace  spanned by the orthogonal 
vectors  $\{ \ket{\chi_n}\}_{n<N_{ph}}$, 
\begin{eqnarray}
\label{eq:dxpcom}
[\tilde{X}, \tilde{P} ]=i,
\end{eqnarray}
and the algebra generated by  $\tilde{X}$ and  $\tilde{P}$ is isomorphic with the algebra generated by $X$ and $P$
on the harmonic oscillator subspace spanned by the HG functions $\{ \ket{\phi_n}\}_{n<N_{ph}}$. 
One can also define  annihilation and  creation operators  on $\dH$  as
\begin{eqnarray}
\label{eq:db}
\tilde{b}^{\dagger}=\frac{1}{\sqrt{2}}\left( \tilde{X} - i \tilde{P}  \right), ~ \tilde{b}=\frac{1}{\sqrt{2}}\left( \tilde{X} + i \tilde{P}  \right),
\end{eqnarray} 
\noindent which satisfy 
\begin{eqnarray}
\label{eq:bdb}
[\tilde{b}, \tilde{b}^{\dagger}]= 1
\end{eqnarray}
\noindent on the subspace $\{ \ket{\chi_n}\}_{n<N_{ph}}$.

On the  subspace $\{ \ket{\chi_n}\}_{n<N_{ph}}$
the discrete  Hamiltonian  
\begin{eqnarray}
\label{eq:tilde_ham}
\H_h=\frac{1}{2} \P2+ \frac{1}{2} \X2,
\end{eqnarray}
\noindent corresponds to the harmonic oscillator Hamiltonian (\ref{eq:hos_ham}).
Therefore, $\{ \ket{\chi_n}\}_{n<N_{ph}}$  are eigenvectors
of $\H_h$ with the eigenspectrum  $\H_h\ket{\chi_n}= \left(n+1/2\right)\ket{\chi_n}$. Moreover, 
$\{ \ket{\chi_n}\}_{n<N_{ph}}$ 
span the low-energy subspace of $\dH$, 
which we will demonstrate by numerically calculating the $N_x$ eigenvalues and eigenvectors of  $\H_h$. 

\subsection{Numerical investigation of the discrete space}
\label{ssec:dnho}

\begin{figure}
\begin{center}
\includegraphics*[width=3.3in]{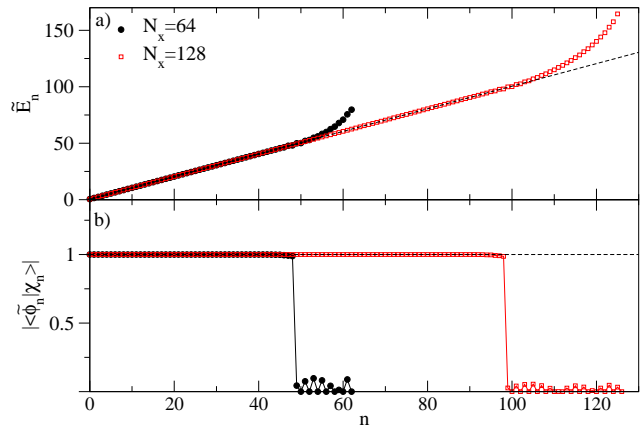}
\caption{
(a)  Energy spectrum $\tilde{E}_n$ of the discrete Hamiltonian $\H_h$ (\ref{eq:tilde_ham}) for $N_x=64$ 
and $N_x=128$. The dashed line is the continuous harmonic oscillator spectrum, Eq.~(\ref{eq:hos_en}).
(b) Overlap between the eigenvectors  $\ket{\tphi_n}$ of $\H_h$,   
and the HG functions projected on the $\{\ket{x_i}\}$ subspace, $\ket{\chi_n}$ (Eq.~(\ref{eq:dhg})).
For $n<N_{ph}$ where $N_{ph}$ is a cutoff number increasing with increasing $N_x$,
$\tilde{E}_n \approx n+\frac{1}{2}$ and $\ket{\tphi_n} \approx \ket{\chi_n}$.}
\label{fig:eigensHd}
\end{center}
\end{figure}

The energy spectrum $\tilde{E}_n$ of $\H_h$ (\ref{eq:tilde_ham}) calculated by exact diagonalization
is shown in Fig.~\ref{fig:eigensHd}(a) for two cases, $N_x=64$ and $N_x=128$, respectively. 
A cutoff number  $N_{ph}$ can be defined such that
the first $N_{ph}$ energy levels are, within a small error, close  to 
the ones corresponding to harmonic oscillator 
energy levels, {\em i.e.,} 
$\tilde{E}_n=E_n+\epsilon$. We will show later in this section that the error $\epsilon$
decreases exponentially by increasing $N_x$ or by decreasing $N_{ph}$.

The low energy eigenstates $\{ \ket{\tphi_n}\}_{n<N_{ph}}$ of $\H_h$ are
the projected HG functions on the discrete basis $\{\ket{\chi_{n}}\}_{n<N_{ph}}$, Eq.~(\ref{eq:dhg}),
in agreement with the theoretical arguments discussed in Section~\ref{ssec:dho}. This can be inferred from Fig.~\ref{fig:eigensHd}(b),
where we see that the overlap $| \bracket{\tphi_{n}}{\chi_n} | =1 - \epsilon$ for $n<N_{ph}$. 
The eigenstates $\{ \ket{\tphi_n} \}$ are calculated by exact diagonalization.
Unlike the low energy states  characterized by large probability at small $x_i$ and exponentially small probability density
at the grid edge ({\em i.e.}, when $x_i \approx \pm L $), 
the eigenstates in the high energy sector,  $\{ \ket{\tphi_n} \}_{N_{ph} \le n<N_x}$, have small probability density
at small $x_i$ and large probability density
close to the  grid edge (not shown). They have a small overlap with the projected HG functions, as Fig.~\ref{fig:eigensHd}(b) shows.

\begin{figure}
\begin{center}
\includegraphics*[width=3.3in]{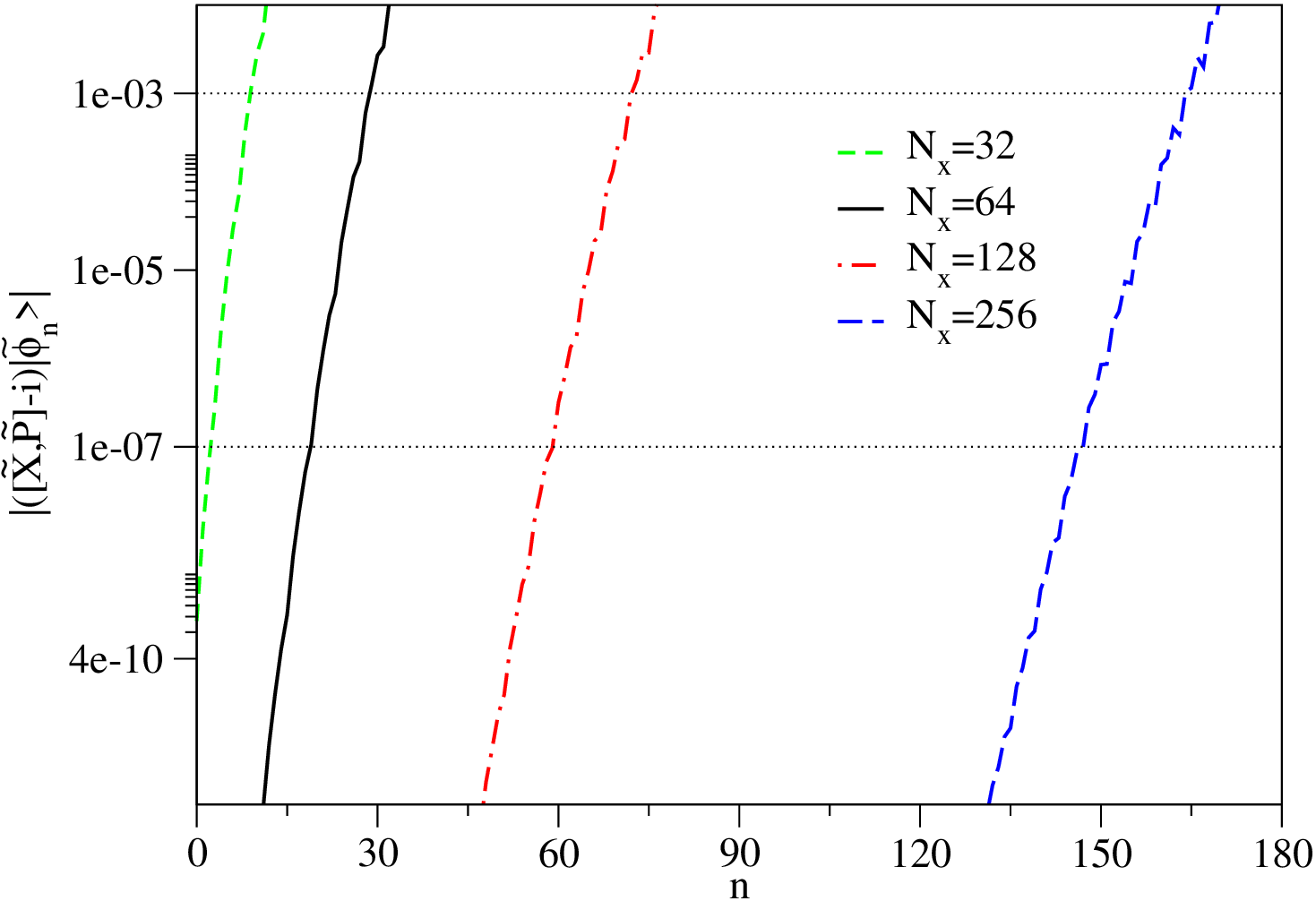}
\caption{$| \left([\tilde{X}, \tilde{P} ] -i \right) \ket{\tphi_n}|$ versus $n$ for different values of $N_x$.
For  ${n<N_{ph}}$ the commutation operator satisfies 
$| \left([\tilde{X}, \tilde{P} ] -i \right) \ket{\tphi_n}|<\epsilon$,
with $\epsilon  \lesssim 10 \exp[-\left(0.51 N_x -0.765 N_{ph} \right)]$.
Up to an exponentially small error, the algebra generated by $\{ \tilde{X}, \tilde{P} \} $
on the low-energy subspace $\{\ket{\tphi_n}\}_{n<N_{ph}}$ of $\dH$ is isomorphic with the 
algebra generated by $\{ X, P \} $ on
low-energy subspace of the harmonic oscillator $\{\ket{\phi_n}\}_{n<N_{ph}}$.
}
\label{fig:xpcomm}
\end{center}
\end{figure}

Fig.~\ref{fig:xpcomm} shows that $|([\tilde{X}, \tilde{P}]-i ) \ket{\tphi_n} | < \epsilon$
for $n < N_{ph}$. The value of $\epsilon$ is exponentially small and it is a consequence of
cutting the tails of the  HG functions for $| x |, | p |>L$. Numerically, we find the convergence rate to be 
\begin{equation}
\label{eq:error}
\epsilon  \lesssim 10 e^{-\left(0.51 N_x -0.765 N_{ph} \right)}.
\end{equation}
 
We conclude that the numerical calculations agree with the analytical predictions, 
supporting the isomorphism  between the $\{ \tilde{X}, \tilde{P} \} $ 
and the $\{X, P \}$ generated algebras  on the low-energy subspace defined by $n < N_{ph}$.

\begin{figure}
\begin{center}
\includegraphics*[width=3.3in]{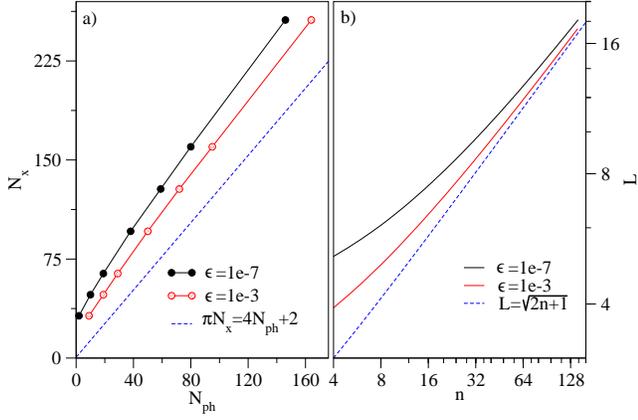}
\caption{
(a) The size of the discrete space, $N_x$, increases linearly with the size of the low-energy subspace, $N_{ph}$.
The full (open) symbols are extracted from Fig.~\ref{fig:xpcomm} for $\epsilon=10^{-7}$ ($\epsilon=10^{-3}$). 
The dashed line is the limit given by the WKB  turning points, Eq.~(\ref{eq:NxNph}).
(b) (Logarithmic scale) $L$, the half-width of the HG functions $\phi_n$, defined by $1-\int_{-L}^{L} | \phi_n(x) |^2 dx=\epsilon$
as a function of $n$. $L$ scales approximately as $\sqrt{n}$.
The dashed line corresponds to the WKB  equation for the turning points, $L=\sqrt{2n+1}$. }
\label{fig:nxnphl}
\end{center}
\end{figure}

The size  $N_x$ of the discrete Hilbert space $\dH$ required 
to accommodate  $N_{ph}$ low  energy states increases approximately linearly with increasing $N_{ph}$.
For example, in Fig.~\ref{fig:nxnphl}(a) we plot the minimum $N_x$ necessary to  have  $N_{ph}$ states
in the low-energy regime with $\epsilon=10^{-7}$  and $\epsilon=10^{-3}$ accuracy. 
The proportionality between $N_x$ and  $N_{ph}$ can be understood by noticing
that the exponential convergence occurs  when  the grid points  $x_i$  cover the
width $2L_{N_{ph}}$ of $\phi_{N_{ph}}(x)$. As shown in Fig.~\ref{fig:nxnphl}(b),
to a first approximation, $L_{N_{ph}} \propto  \sqrt{N_{ph}}$, which is not surprising taking into account that
the turning points in the WKB approximation are defined by  $\sqrt{2N_{ph}+1}$ (see the discussion before Eq.~(\ref{eq:NxNph})).
On the other hand, 
the width covered by  $N_x$ points is $\sqrt{2 \pi N_x}$, thus being proportional to $\sqrt{N_x}$ and
implying linear dependence of $N_x(N_{ph})$.

\subsection{Cutoff of the maximum boson occupation number}
\label{ssec:nph}

As long as the physical problem of interest can be addressed by truncating the number of bosons
per state our representation is suitable for quantum computation. 
In most cases the boson distribution number is 
Poissonian,  falling exponentially fast to zero with increasing the number of bosons.
For electron-phonon systems the cutoff on the maximum phonon occupation number depends on the effective strength of the
interaction, on the size of the low-energy space under consideration and on the desired precision, as discussed below.

In order to understand the  truncation of the boson space, let us  focus on
a particular harmonic oscillator.  
The states belonging to the chosen harmonic oscillator space evolve under the action of a
forced harmonic oscillator Hamiltonian, {\em i.e.}, a harmonic oscillator  with a  displacing force. 
The effective force is determined by the configuration of the fermions and the bosons coupling to
the oscillator.

For example, let us consider the electron-phonon system, Eq.~(\ref{eq:gen_elham}), and focus on
the harmonic oscillator labeled $n\nu$.  The Hamiltonian can be written  as
\begin{eqnarray}
\label{eq:Hred}
H=H_{n\nu} +F X_{n\nu} + B X_{n\nu}+H_1 
\end{eqnarray}
\noindent where 
\begin{eqnarray}
\label{eq:hnnu}
H_{n\nu} &=&\frac{P^2_{n\nu}}{2 M_{\nu} } + \frac{1}{2} M_{\nu}\omega^2_{n \nu} X^2_{n\nu}  \\
\label{eq:fx}
F X_{n\nu}&=&\left[\sum_{ij  } g_{ijn\nu} \left( c^{\dagger}_i c_j +c^{\dagger}_j c_i  \right) \right] X_{n\nu} \\
\label{eq:bx}
B X_{n\nu}&=&\left[\sum_{m\mu \ne n \nu}K_{m\mu n\nu} X_{m\mu} \right] X_{n\nu}.
\end{eqnarray}
\noindent $H_1$ contains the remaining terms in the Hamiltonian which do not act on the Hilbert space of the 
harmonic oscillator  $n\nu$.
Let us consider an arbitrary state $\ket{\Phi}$ which we write as
\begin{eqnarray}
\label{eq:phigen}
\ket{\Phi}=\sum_{\alpha \beta s} c_{ \alpha \beta s} \ket{f_\alpha} \otimes \ket{b_\beta}  \otimes \ket{s}_{n\nu},
\end{eqnarray}
\noindent where $\ket{f_\alpha}\otimes \ket{b_\beta}  \otimes \ket{s}_{n\nu}$ form a complete basis set.
 In Eq.~(\ref{eq:phigen}) the vectors $\ket{f_\alpha}$ span the electron Hilbert space while the vectors
 $\ket{b_\beta}  \otimes \ket{s}_{n \nu}$ span the full  phonon space. The vectors $\ket{s}_{n\nu}$
 belong to the Hilbert space of the harmonic oscillator  $n\nu$.
 We choose the vectors $\ket{f_\alpha}$ such that they are  eigenvectors of $F$, {\em i.e.,} 
 $F\ket{f_\alpha}=f_\alpha\ket{f_\alpha}$, and the vectors $\ket{b_\beta}$ such that they are  eigenvectors of $B$, {\em i.e.,} 
 $B\ket{b_\beta}=b_\beta \ket{b_\beta}$. Let us now imagine 
 a path integral  or Trotter-Suzuki expansion~\cite{trotter_1959, suzuki_1976} of the evolution operator
 in small time steps. 
The evolution operator for a small time step $\t$ is
\begin{eqnarray}
\label{eq:ev_f}
&&e^{-i\t H}\ket{\Phi}  \approx  e^{-i \t H_1}  e^{-i \t (H_{n\nu}+ FX_{n\nu} +BX_{n\nu}) }\ket{\Phi}\\ \nonumber 
&&=\sum_{\alpha \beta s} c_{\alpha \beta s}  \left( e^{-i \t H_1} \ket{f_\alpha} \otimes \ket{b_\beta} \right) 
\otimes \\ \nonumber 
&&\left( e^{-i \t \left( H_{n\nu}+\left( f_\alpha +b_\beta \right) X_{n\nu}  \right) } \ket{s}_{n\nu} \right). 
\end{eqnarray}
\noindent From Eq.~(\ref{eq:ev_f}) one can see that the
evolution operator of a full electron-phonon system
implies a superposition of forced harmonic Hamiltonians, $H_{n \nu}+g_{\alpha \beta} X_{n\nu}$ acting on the phonon space $n\nu$ at every time step. 
The coupling strength at that particular time step, $g_{\alpha \beta}=f_\alpha + b_\beta$, 
depends on the 
configuration  $\alpha$ of the electrons interacting with the harmonic oscillator $n \nu$
and the configuration $\beta$ of the  phonons interacting with the harmonic oscillator $n \nu$.

The forced harmonic oscillator problem can be solved 
exactly~\cite{merzbacher}. (See also Appendix~\ref{app:fho}.)
It turns out that the low-energy space of the forced harmonic oscillator
can be obtained from the low energy space of the unperturbed
harmonic oscillator via  displacement operators.

Let us assume first that the coupling $g_{\alpha \beta}$ is constant. In that case
the forced harmonic oscillator is just a harmonic oscillator 
with a displaced equilibrium position,
\begin{eqnarray}
\label{eq:hos_ex}
H_g&=&\frac{P^2}{2M}+\frac{1}{2}M \omega^2X^2+gX\\ \nonumber
&=&\frac{P^2}{2M}+\frac{1}{2}M \omega^2 (X + \frac{g}{M \omega^2})^2-\frac{g^2}{M\omega^2}.
\end{eqnarray}
\noindent The term $-g^2/M\omega^2$ is a constant which represents the deformation energy. 
The eigenstates $\{\ket{\phi_{gn}}\}$ of $H_g$ are obtained by  applying the displacement operator (see Eq.~(\ref{eq:dis})),  
\begin{eqnarray}
\label{eq:dis_p}
e^{i P \frac{g}{M \omega^2}}=D(-\frac{g}{\omega \sqrt{2M\omega}})=e^{-\frac{g}{\omega \sqrt{2M\omega} } (b^{\dagger}-b)}
\end{eqnarray}
\noindent on the unperturbed harmonic oscillator $H_0$ eigenstates $\{\ket{\phi_{n}}\}$, 
\begin{eqnarray}
\label{eq:ng_n}
\ket{\phi_{gn}}=D(-\frac{g}{\omega \sqrt{2M\omega}}) \ket{\phi_{n}},
\end{eqnarray}
\noindent {\em i.e.}, $\{\ket{\phi_{gn}}\}$ are {\em displaced number states}, Eq.~(\ref{eq:dzn}).

In general $g_{\alpha \beta}$ is not  constant, since it depends on the configuration of the
surrounding fermions and  bosons interacting with the oscillator, and these  surroundings changes at every time step. 
The evolution of a displaced number state $\ket{n,z} \equiv D(z)\ket{\phi_{n}}$ under the action of 
the forced harmonic oscillator Hamiltonian with time dependent force $g_{\alpha \beta}(t)$ 
is given by (see Eqs.~(\ref{eq:fho_ev}) and (\ref{eq:app_ugnz})) 
\begin{equation}
\label{eq:ugnz}
U(t)\ket{n,z}= e^{i \gamma} e^{i\left(\delta-n \omega t\right)} \ket{n, \left(\zeta_{\alpha \beta}(t) +z\right)e^{-i \omega t}}.
\end{equation}
\noindent where $\gamma$ and $\delta$ are real phases and $\zeta_{\alpha \beta}(t)$ is 
\begin{eqnarray}
\label{eq:zeta}
\zeta_{\alpha \beta}(t)&=& -\frac{i}{\sqrt{2M\omega}} \int^t_0 g_{\alpha \beta}(u) e^{i \omega u} du.
\end{eqnarray}
\noindent 
Thus, at any time the evolution of a  displaced number state is a  displaced number state.
Therefore,  an initial low energy state written
as a linear combination of  displaced number states
will remain  a linear combination of  displaced number states.

As discussed in Appendix~\ref{app:fho}, the boson occupation number of a displaced number state 
is Poissonian and  falls exponentially to zero with increasing boson number.
As long as the displacements are bounded
one can chose a cutoff value $N_{ph}$ for truncating the boson Hilbert space to the desired accuracy.
Equation~(\ref{eq:nphscaling}) implies that
\begin{eqnarray}
\label{eq:scnphe}
N_{ph}&=&\O\left( \sqrt{\ln(\epsilon^{-1})} \right),\\
\label{eq:scnphnzeta}
N_{ph}&=&\O\left(|\zeta_{max}|^2\right),\\
\label{eq:scnphne}
N_{ph}&=&\O\left(N_E\right),
\end{eqnarray}
\noindent where $\epsilon$ is the accuracy,  $N_E$ 
is the size of the low energy space under consideration and $|\zeta_{max}| = \max_{t, \alpha, \beta}|\zeta_{\alpha, \beta}(t)|$  is the maximum displacement 
of the harmonic oscillator under the action of the effective force.  
Since the maximum displacement is proportional to the coupling $g_{\alpha \beta}$ (see Eq.~(\ref{eq:zeta})), one can regard  $|\zeta_{max}|^2 $ as the 
effective coupling strength between the harmonic 
oscillator and the environment.

Note that the displacement is not  bounded if the  system is unstable 
(and $g_{\alpha \beta}$ is unbounded)
or when $g_{\alpha \beta}(t)$ is resonant  with the oscillator and  $\zeta_{\alpha \beta}(t)$ grows linearly with time (see Eq.~(\ref{eq:zeta})).

\section{Algorithm}
\label{sec:algor}

Our algorithm  simulates  the evolution operator $e^{-i t H }$ of the fermion-boson systems
on a gate quantum computer.
As in the fermion algorithms, we employ the Trotter-Suzuki expansion~\cite{trotter_1959, suzuki_1976} of the evolution operator 
to a product of  short-time evolution operators corresponding to the
noncommuting terms in the Hamiltonian. For a Hamiltonian  written as  $H=\sum_m H_m$,  the first-order Trotter decomposition reads
\begin{eqnarray}
\label{eq:trotter}
e^{-i H \Delta t} = \prod_m e^{-i H_m \Delta t} + {\cal{O}} (\Delta t^2).
\end{eqnarray}
\noindent Higher-order decomposition schemes can be used for optimal performance, but that type of optimization is not 
addressed in this paper. Our main focus is presenting  algorithmic circuits for the small time  evolution operators 
corresponding to the terms in the fermion-boson Hamiltonian.

While the polaron example
presented in Section~\ref{sec:polaron} below is based on the QPE method, our  algorithm 
can be utilized using other techniques such as the variational quantum eigensolver   (VQE)~\cite{peruzzo_nature_2014,mcclean_NJPhys_2016} and 
adiabatic state preparation~\cite{Farhi_science_2001}.

\subsection{Qubit representation of harmonic oscillator space}
\label{ssec:qb_rep}

On a gate quantum computer each harmonic oscillator state is represented
as a superposition of  $N_x=2^{n_x}$ discrete states $\{\ket{x_j}\}$ 
and stored in a register of $n_x$ qubits, 
\begin{eqnarray}
\label{eq:phi_gen}
\ket{\phi}=\sum_{j=0}^{2^{n_x}-1} \phi_j \ket{x_j}.
\end{eqnarray}
\noindent  The operators $X$ and $P$ acting on the boson space are 
replaced by their discrete 
versions $\tilde{X}$ (Eq.~(\ref{eq:tilde_x})) and $\tilde{P}$ (Eq.~(\ref{eq:tilde_p})), respectively. 
The states $\{\ket{x_j}\}$ are eigenvectors of $\tilde{X}$. 
Since the number stored in the register $\ket{x_j}$ in Eq.~(\ref{eq:phi_gen}) is 
between $0$ and $N_x-1$, the eigenvalues $x_j$ 
\begin{equation}
\label{eq:Xeigenqubit}
\tilde{X} \ket{x_j}=x_j\ket{x_j},
\end{equation}
\noindent are 
\begin{equation}
\label{eq:Xqubit}
x_j=(j-N_x/2)\Delta,~~j=\overline{0, N_x-1}.
\end{equation}
The states $\{\ket{p_n}\}$ obtained via discrete Fourier transform (or  quantum Fourier transform) from the states $\{\ket{x_j}\}$, 
\begin{equation}
\label{eq:FTqubit}
\ket{p_n} = \frac{1}{\sqrt{N_x}}\sum_{j=0}^{ N_x-1 } e^{i \frac{2 \pi}{N_x} j n} \ket{x_j},~~n=\overline{0, N_x-1},
\end{equation}
\noindent are eigenvectors of  $\tilde{P}$. Nevertheless the states $\{\ket{p_n}\}$ in 
Eq.~(\ref{eq:FTqubit}) are different from the ones defined by Eq.~(\ref{eq:oFFT_xp}),
since in Eq.~(\ref{eq:FTqubit}) the Fourier transform is not centered.
As a consequence   one has
\begin{equation}
\label{eq:Peigenqubit}
\tilde{P}=p_n\ket{p_n},
\end{equation}
\noindent with
\begin{eqnarray}
\label{eq:Pqubit}
p_n&=& n\Delta , ~~n=\overline{0, N_x/2-1}\\
p_n&=&  (n-N_x)\Delta , ~~n=\overline{N_x/2, N_x-1}.
\end{eqnarray}

\subsection{Noninteracting Boson Hamiltonian}
\label{ssec:ev_ph}

\begin{figure*}[tb]
\begin{center}
\includegraphics*[width=4.4in]{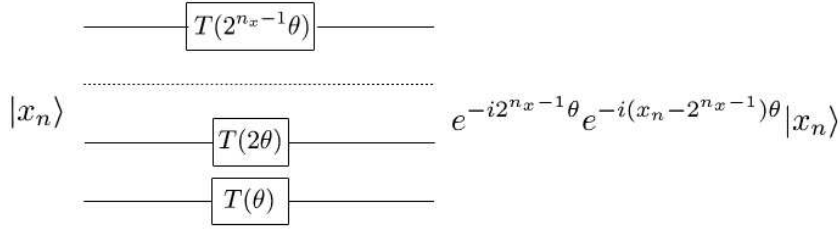}
\caption{The circuit for $e^{-i \t \tilde{X}_n}  \ket{x_n}$.
On each qubit a phase shift gate is applied. The angles of the phase shift gates are determined by writing  
$x_n=\sum^{n_x-1}_{r=0} x^r_n 2^r$,  where $\{x^r_n\}_{r=\overline{0,n_x-1}}$ takes binary values. 
A phase factor $\exp(-i 2^{n_x-1} \t)$
accumulates at every Trotter step.}
\label{fig:ps_x}
\end{center}
\end{figure*}

In this section we discuss the implementation of the evolution operators corresponding to 
the different terms in the noninteracting boson Hamiltonian (\ref{eq:xp_b0ham}).
Unlike the notation used in Section~\ref{ssec:qb_rep} where the index $j$ in $\ket{x_j}$ (or $n$ in $\ket{p_n}$)
was used to label the basis states of a single harmonic oscillator, here and in the following sections  
a vector $\ket{x_n}$ represents a state of the harmonic oscillator $n$ ($n$ is a site label).
 
A circuit for the term  
\begin{eqnarray}
\label{eq:ev_x}
e^{-i \t \tilde{X}_n} \ket{x_n},
\end{eqnarray}
\noindent  is shown in Fig.~\ref{fig:ps_x}. The factor $\Delta$ (see Eq.~\ref{eq:Xqubit}) is absorbed into 
the definition of $\t$.
On every qubit  belonging to the boson register $\ket{x_n}$ a controlled phase shift gate 
\begin{eqnarray}
\label{eq:Tgate}
T(\t_r)=
\begin{bmatrix}
1 & 0\\
0 & e^{-i \t_r}
\end{bmatrix},
\end{eqnarray}
\noindent with $\t_r=2^{r}\t$ is applied. Here $r$ represents the qubit index defined by the
binary representation of $x_n$, {\em i.e.}, 
$x_n=\sum^{n_x-1}_{r=0} x^r_n 2^r$, with $\{x^r_n\}_{r=\overline{0,n_x-1}}$ taking binary values.
The $N_x/2$ term entering in  Eq.~\ref{eq:Xqubit} yields
a phase factor equal to $\exp(-i 2^{n_x-1} \t)$  which 
accumulates to the wave function at each Trotter step.  This phase factor can be tracked classically.

\begin{figure*}[tb]
\begin{center}
\includegraphics*[width=6in]{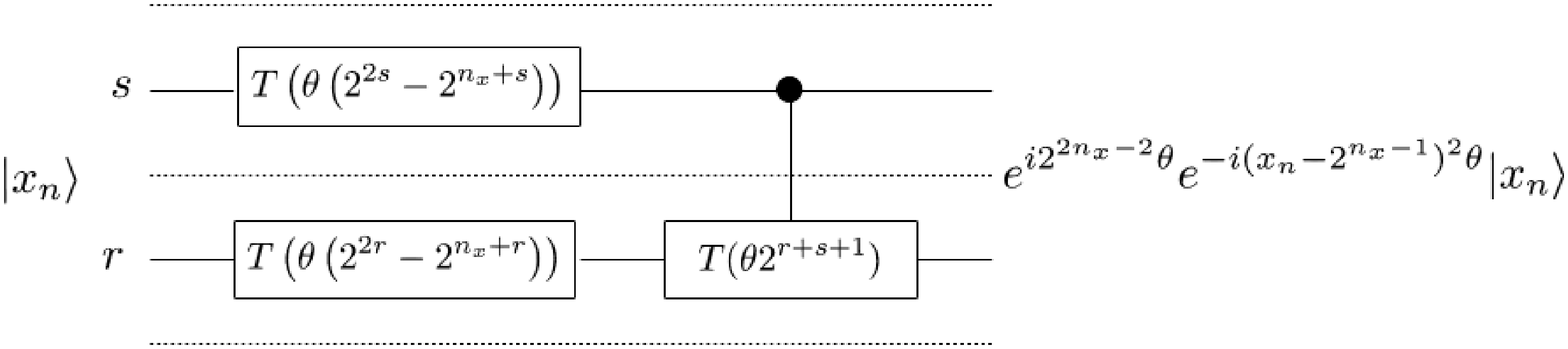}
\caption{ The circuit for $e^{-i \t \X2_n} \ket{x_n}$ requires $n_x$ phase shift gates (one on each qubit) and $n_x (n_x-1)/2$  controlled  phase shift gates.
The angles of the phase shift gates are determined by writing 
$(x_n-2^{n_x-1})^2=\sum^{n_x-1}_{r=0} x_n^r \left( 2^{2r}-2^{n_x+r} \right) +\sum_{r<s} x_n^r x_n^s 2^{r+s+1}+2^{2 n_x-2}$, 
where $\{x_n^r\}_{r=\overline{0,n_x-1}}$ is the binary representation of $x_n$, {\em i.e.,} $x_n=\sum^{n_x-1}_{r=0} x_n^r 2^{r}$. 
A phase factor $\exp(i 2^{n_x-2} \t)$ accumulates at every Trotter step.}
\label{fig:ps_x2}
\end{center}
\end{figure*}

The implementation of
\begin{eqnarray}
\label{eq:ev_x2}
e^{-i \t \X2_n} \ket{x_n} 
\end{eqnarray}
\noindent requires phase shift gates and is shown in Fig.~\ref{fig:ps_x2}.
Note that, unlike the circuit for Eq.~(\ref{eq:ev_x}), the circuit for Eq.~(\ref{eq:ev_x2}) requires extra
$n_x (n_x-1)/2$  controlled  phase shift gates, a consequence of squaring  $x_n$.
The angles for the phase shift gates are 
determined by  writing $(x_n-N_x/2)^2$ in the binary format (see the figure's caption).

\begin{figure*}[tb]
\begin{center}
\includegraphics*[width=5in]{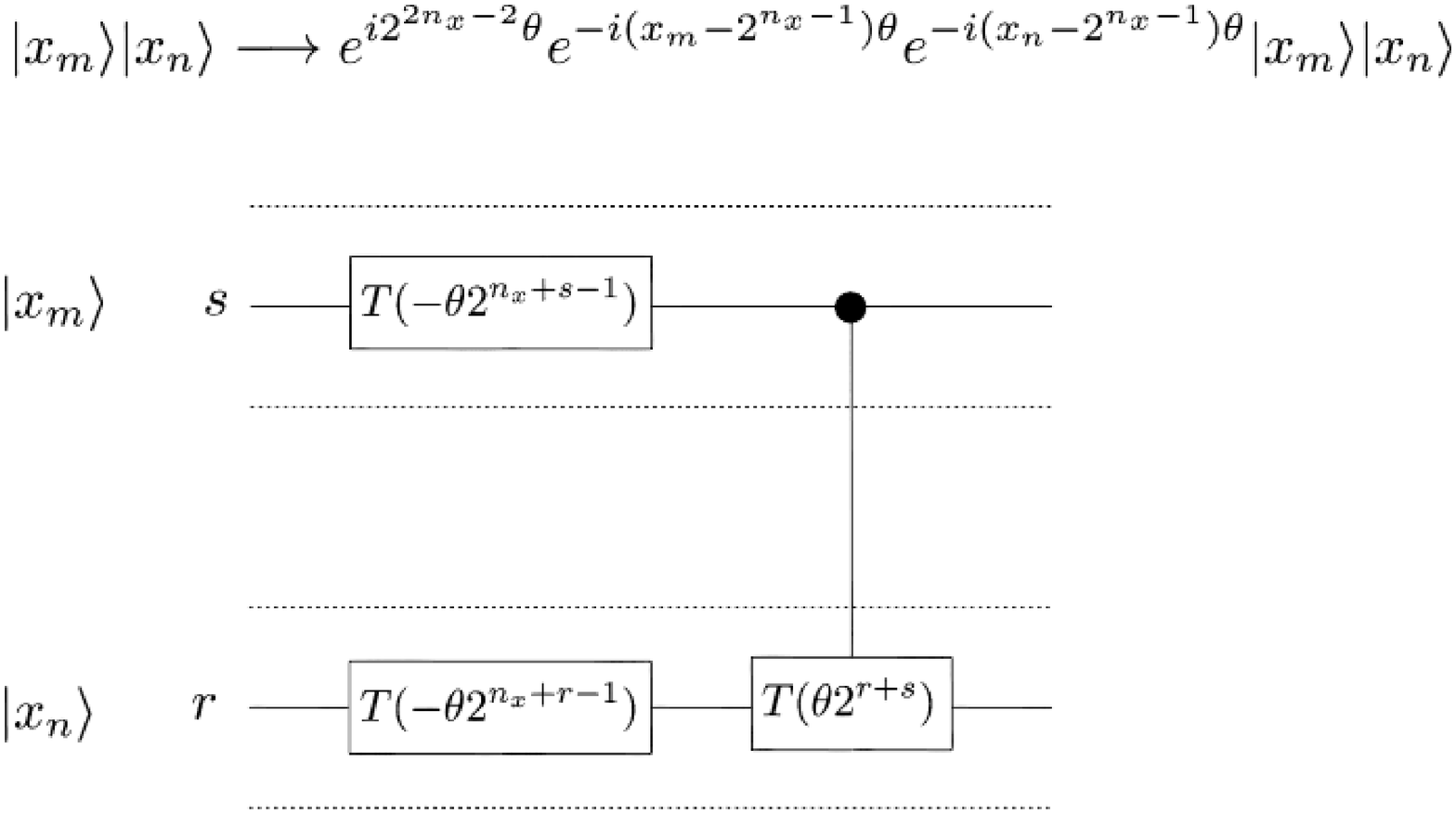}
\caption{ The circuit for $e^{-i \t \tilde{X}_n \tilde{X}_m} \ket{x_n}\ket{x_m}$ 
requires $2 n_x$ phase shift gates (one on each qubit of the two boson registers)
and $n_x^2$  controlled  phase shift gates.
The angles of the phase shift gates are determined by writing 
$(x_n-2^{n_x-1}) (x_m-2^{n_x-1})=\sum^{n_x-1}_{r,s=0}x_n^r x_m^s 2^{r+s}-\sum^{n_x-1}_{r=0}(x_n^r +x_m^r) 2^{r+n_x-1}+2^{2 n_x-2}$,
where $\{x_n^r\}_{r=\overline{0,n_x-1}}$ is the binary representation of $x_n$, {\em i.e.,} $x_n=\sum^{n_x-1}_{r=0} x_n^r 2^{r}$. 
A phase factor $\exp(i 2^{n_x-2} \t)$ accumulates at every Trotter step.}
\label{fig:ps_KXX}
\end{center}
\end{figure*}

The evolution operator 
\begin{eqnarray}
\label{eq:ev_kxx}
e^{-i \t \tilde{X}_n \tilde{X}_m}\ket{x_n}\ket{x_m} 
\end{eqnarray}
\noindent describes the coupling between oscillators $n$ and $m$ 
and requires two boson registers, as shown in Fig.~\ref{fig:ps_KXX}.
The circuit is similar to the ones for Eqs.~(\ref{eq:ev_x}) and (\ref{eq:ev_x2}),
consisting in phase shift gates.
The phase shift angles are determined by writing the product $(x_n-N_x/2)(x_m-N_x/2)$ as a sum with binary coefficients 
(see the figure's caption). The circuit reduces to $n^2_x$ controlled phase shift gates and $2 n_x$  phase shift gates.

For the implementation of $e^{-i \t \tilde{P}_n} \ket{x_n}$, $e^{-i \t \P2_n} \ket{x_n}$
and $e^{-i \t \tilde{P}_n \tilde{X}_m} \ket{x_n} \ket{x_m} $ ($n \ne m$),
one first applies a  quantum Fourier transform (QFT)~\cite{nielsen_2010}
$\ket{x_n}\xrightarrow{QFT}  \ket{p_n}$.
Then  $e^{-i \t \tilde{P}_n}\ket{p_n}$,  
$e^{-i \t \P2_n}\ket{p_n}$ and 
$e^{-i \t \tilde{P}_n \tilde{X}_m} \ket{p_n} \ket{x_m} $
are implemented by  circuits similar to the ones shown in Fig.~\ref{fig:ps_x}, Fig.~\ref{fig:ps_x2},
and Fig.~\ref{fig:ps_KXX}, respectively. 
These circuits contain phase shift gates with angles determined by writing the eigenvalues
of the operators $\tilde{P}_n$, $\P2_n$ and $\tilde{P}_n \tilde{X}_m$ 
in binary representation.
The last step is
an inverse QFT $\ket{p_n}\xrightarrow{IQFT}  \ket{x_n}$. The idea of implementing
the Hamiltonian terms which are functions of the momentum operator by going to the momentum basis and back  
via Fourier transform was first discussed in Refs.~\cite{zalka_1998,wiesner_1996}.

The implementation of $e^{-i \t \tilde{P}_n \tilde{P}_m} \ket{x_n} \ket{x_m}$
requires two  QFT transforms of the $\ket{x_n}$ and $\ket{x_m}$ registers, such that
$\ket{x_n} \ket{x_m}\xrightarrow{QFT}  \ket{p_n} \ket{p_m}$. The operator 
$e^{-i \t \tilde{P}_n \tilde{P}_m} \ket{p_n} \ket{p_m}$
is implemented in an analogous way to the one shown in Fig.~\ref{fig:ps_KXX}. The circuit ends with 
two  inverse QFT transforms $ \ket{p_n} \ket{p_m} \xrightarrow{IQFT} \ket{x_n} \ket{x_m}$.

\subsection{Fermion Hamiltonian}
\label{ssec:el_ev}
The algorithm for fermions is described at length in numerous papers (see, for example, Refs.~\cite{somma_gubernatis_2002, whitfield_2011,troyer_pra_2015}.)
We assume here an implementation  which  requires  Jordan-Wigner mapping of the fermion operators to
the Pauli operators $X$, $Y$, and $Z$ as in Ref~\cite{troyer_pra_2015}.
Each fermion orbital requires a qubit. The qubit state $\ket{\uparrow} \equiv \ket{0}$ corresponds 
to an unoccupied fermion orbital, while the qubit state $\ket{\downarrow} \equiv \ket{1}$ corresponds to an occupied orbital.

\subsection{Fermion-boson interaction Hamiltonian}
\label{ssec:elph_ev}

\begin{figure*}[tb]
\begin{center}
\includegraphics*[width=5in]{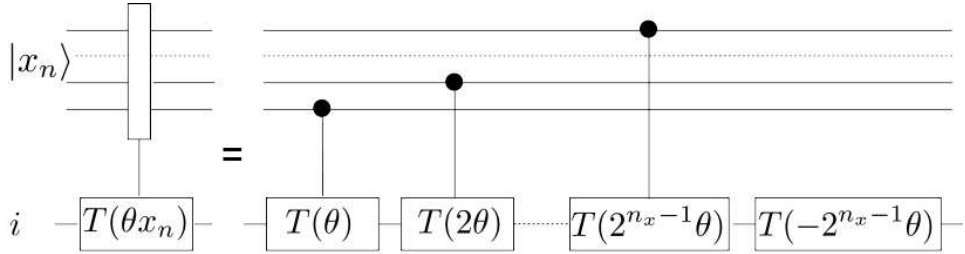}
\caption{ Circuit describing  $e^{-i \t c^{\dagger}_i c_i \tilde{X}_n} \ket{i}  \otimes \ket{x_n}$.
The phase shift angle is proportional to the boson coordinate,
 $x_n-2^{n_x-1}=\sum^{n_x-1}_{r=0} x^r_n 2^r-2^{n_x-1}$, where $\{x^r_n\}_{r=\overline{0,n_x-1}}$ take binary values.}
\label{fig:pase_sf_x}
\end{center}
\end{figure*}

The interaction Hamiltonian acts on both the fermion and boson parts of the Hilbert space
and involves coupling between  single-particle fermion operators and the boson $X$ or $P$ operators,
as described by  Eqs.~(\ref{eq:fbham1}), (\ref{eq:fbham2}), (\ref{eq:fbham3}) and (\ref{eq:fbham4}).

The expressions for the single-particle fermion operators  as  functions of the Pauli operators are
\begin{eqnarray}
\label{eq:cici}
n_i &=&c^{\dagger}_i c_i= \frac{1-Z_i}{2}, \\ 
\label{eq:cicj}
c^{\dagger}_i c_j+c^{\dagger}_j c_i &=& 
\frac{1}{2}\left(X_i X_j+Y_i Y_j \right) Z_{i+1}...Z_{j-1},  \\
\label{eq:jcicj}
i\left(c^{\dagger}_i c_j-c^{\dagger}_j c_i\right) &=& \frac{1}{2} \left(Y_i X_j-X_i Y_j \right) Z_{i+1}...Z_{j-1},
\end{eqnarray}
\noindent where we assume $j>i$.
The implementation of the corresponding  evolution operators for these pure fermion operators 
requires circuits with phase shift $T(\t)$, Eq.~(\ref{eq:Tgate}), or z-rotation 
\begin{eqnarray}
\label{eq:Rzgate}
R_z(\t)=
\begin{bmatrix}
e^{i \frac{\t}{2}} & 0\\
0 & e^{-i \frac{\t}{2}}
\end{bmatrix},
\end{eqnarray}
\noindent gates~\cite{whitfield_2011,troyer_pra_2015}.

The implementation of the fermion-boson interaction  is similar. 
In the case of fermions coupling with the boson position operator $X$ (Eqs.~(\ref{eq:fbham1})  and (\ref{eq:fbham3}))
the only difference is the rotation angle $\t$, which is replaced by $\t x$, where $x$ is the eigenvalue of $\tilde{X}$ corresponding to the boson state $\ket{x}$.

For example, in  Fig.~\ref{fig:pase_sf_x} we show the implementation of 
\begin{eqnarray}
\label{eq:ev_Hep_ciiX}
e^{-i \t c^{\dagger}_i c_i \tilde{X}_n} \ket{i} \otimes \ket{x_n}=\left( T(\theta x_n) \ket{i} \right) \otimes \ket{x_n}
\end{eqnarray}
\noindent where $\ket{i}$ is the  $i$ fermion orbital  and $\ket{x_n}$ is the state of the harmonic oscillator $n$.

\begin{figure*}[tb]
\begin{center}
\includegraphics*[width=5in]{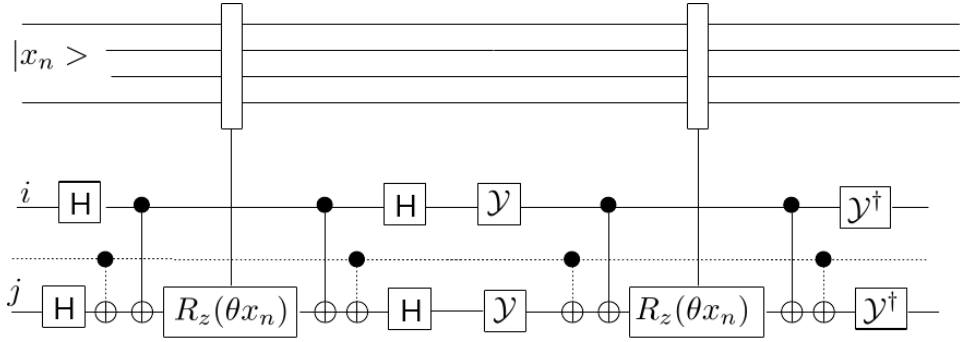}
\caption{The circuit implementing $e^{-i \t \left( c^{\dagger}_i c_j+ c^{\dagger}_j c_i \right) \tilde{X}_n}$
is similar to the circuit for the fermion hopping term (see Fig.~(9) of Ref~\cite{troyer_pra_2015} or Table~A1 of Ref.~\cite{whitfield_2011}).
The angle of  the $z$-rotation gates is  $\t x_n$.}
\label{fig:fb_hopping}
\end{center}
\end{figure*}

The evolution of the term coupling the fermion hopping operator to the boson position operator can be written as 
\begin{eqnarray}
\label{eq:hop_map}
e^{-i \t \left(c^{\dagger}_i c_j +c^{\dagger}_j c_j\right) \tilde{X}_n} 
\approx \Y^{\dagger}_i \Y^{\dagger}_j e^{-i \t Z_{i} Z_{i+1}...Z_{j} \tilde{X}_n} \Y_j \Y_i \\ \nonumber
H_i H_j e^{-i \t Z_{i} Z_{i+1}...Z_{j} \tilde{X}_n} H_j H_i.
\end{eqnarray}
\noindent Since  $H$ (Hadamard) and $\Y$ ($\Y=R_x(\frac{\pi}{2})$) operators
satisfy $H X H=Z$ and $\Y^{\dagger} Y \Y=Z$,
they are  employed
to rotate the Pauli $X$ and $Y$ operators, respectively (see Eq.~(\ref{eq:cicj})), to the  $Z$ operator.
The circuit is shown in Fig.~\ref{fig:fb_hopping}. It is  similar to the circuit shown 
in Fig.~(9) of Ref~\cite{troyer_pra_2015} or Table~A1 of Ref.~\cite{whitfield_2011}
for $e^{-i \t (c^{\dagger}_i c_j +c^{\dagger}_j c_i)}$.  
The difference is that  $R_z(\t)$ is replaced by  $R_z(\t x_n)$.
The circuit for $R_z(\t x_n)$  is similar to the one shown in  Fig.~\ref{fig:pase_sf_x}, the only difference being
that $T$ (the phase shift gate)  is replaced by $R_z$ (the z-rotation gate).

The evolution of the term coupling the fermion current operator (see Eq.~(\ref{eq:cicj}))
to the boson position operator is
\begin{eqnarray}
\label{eq:j_map}
e^{-i \t i \left(c^{\dagger}_i c_j -c^{\dagger}_j c_j\right) \tilde{X}_n} 
\approx H_i \Y^{\dagger}_j e^{i \t Z_{i} Z_{i+1}...Z_{j} \tilde{X}_n} \Y_j H_i \\ \nonumber
\Y^{\dagger}_i H_j e^{-i \t Z_{i} Z_{i+1}...Z_{j} \tilde{X}_n} H_j \Y_i.
\end{eqnarray}
\noindent The circuit is analogous  the one shown in Fig.~\ref{fig:fb_hopping},
but  the order of the single qubit operators acting
on the $i$ fermion orbital is $\Y \Y^{\dagger}HH$ instead of  $HH \Y \Y^{\dagger}$ 
(on the $j$ fermion qubit the order remains the same, $HH \Y \Y^{\dagger}$). The other  difference is
that the  rotation angle corresponding to the second $R_z$ gate changes sign, 
{\em i.e.,} it is $-\t x_n$ instead of $\t x_n$.

For the cases when the fermions couple  to the boson momentum operator $P$
as in Eqs.~(\ref{eq:fbham2})  and (\ref{eq:fbham4}),
the circuits are analogous to the ones 
corresponding to  Eqs.~(\ref{eq:fbham1})  and (\ref{eq:fbham3}),
respectively. The main difference is that the rotation angle acting on the fermion qubits 
is proportional to $\t p$ instead of $\t x$. 
The value $p$ is the eigenvalue of $\tilde{P}$ corresponding to the boson state $\ket{p}$ in the momentum representation.
The state $\ket{p}$ is obtained by applying  QFT to $\ket{x}$ at the beginning of the circuit. An inverse QFT transformation 
should be applied at the end of the circuit from  $\ket{p}$ to
$\ket{x}$ to restore the position representation.

The nonlocality of the Jordan-Wigner mapping 
increases the circuit depth for the fermionic terms in 
the Hamiltonian~\cite{somma_gubernatis_2002, whitfield_2011, troyer_pra_2015}.
In the case when the fermion-boson coupling involves the fermion hopping operator,
the additional contribution to the circuit depth of the Jordan-Wigner strings associated with 
the fermion-boson interaction can be avoided by combining 
the implementation of the  fermion and fermion-boson terms.
For example, for a coupling type characteristic of electron-phonon systems such as in Eq.~(\ref{eq:fbham1}),  one can  implement 
\begin{eqnarray}
\label{eq:ev_Hep_1pcpqX}
e^{-i\left(  c^{\dagger}_i c_j + c^{\dagger}_i c_j   \right) \left(\t_0 +\sum_m \t_m \tilde{X}_m\right) },
\end{eqnarray}
\noindent which reduces to the circuit shown in Fig.~\ref{fig:fb_hopping} with 
$R_z(\t_0+\sum_m \t_m x_m)$ gates replacing the $R_z(\t x_n)$ ones.
In this case the contribution to the circuit depth for long-range fermion-boson interactions is 
$\O(N)$.

\subsection{Squeezing and boson-boson interaction Hamiltonian}
\label{ssec:ev_bosonboson}

As discussed in Section \ref{ssec:bh}, the implementation of the squeezing Hamiltonian (\ref{eq:bsham}) and
the boson-boson interaction Hamiltonian (\ref{eq:b1ham}) reduces to the implementation of terms of
type $A_nA_m$,  $A_{n}A_{m}A_{r}$ and $A_{n} A_{m} A_{r} A_{s}$, where $A_{n}$ is either the
$X_{n}$ or the $P_{n}$ operator of the harmonic oscillator $n$.

The implementation of terms $A_nA_m$ ($n \ne m$) present in the squeezing Hamiltonian~(\ref{eq:xp_bsham})
was described in Section~\ref{ssec:ev_ph}. The implementation of the interaction terms which  do not contain
self-interacting $XP$ products such as  $X^v_nP^u_n$ where  $v,u>0$ integers, is a straightforward
generalization. For example, the corresponding circuit for  $A_{n} A_{m} A_{r} A_{s}$
consists of phase shift, controlled phase shift, double-controlled phase shift and 
triple-controlled phase shift gates applied on the qubits of the boson registers $n$, $m$, $r$ and $s$.
The angles of the phase shift gates are determined by writing $a_{n} a_{m} a_{r} a_{s}$ as a sum  with binary coefficients,
where  $a_n$ is the eigenvalue of the operator $A_{n}$ corresponding to the state of the  boson register $n$.
For  $A_n \equiv P_n$ the circuit starts with a QFT and ends with an inverse QFT on the boson register $n$.

\begin{figure*}[tb]
\begin{center}
\includegraphics*[width=6in]{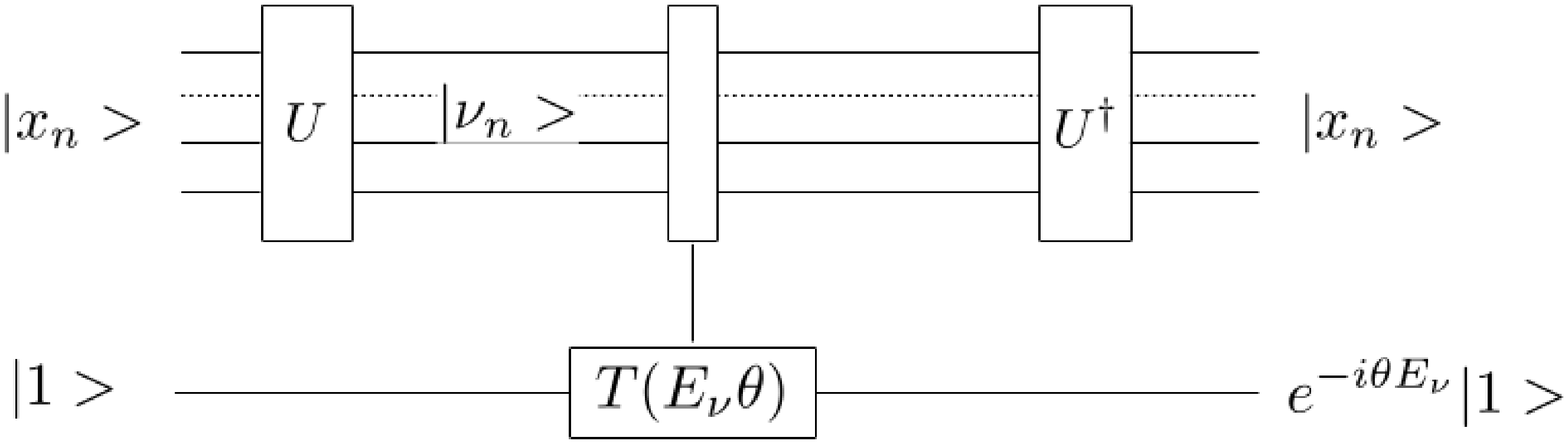}
\caption{ The circuit implementing $e^{-i \t H_{xp}}$. The unitary operator $U$ transforms the position vectors $\ket{x}$
into the eigenvectors $\ket{\nu}$ of $H_{xp}$. For every state $\ket{\nu}$ a phase shift with the angle $E_\nu \t$
is applied to an ancilla qubit prepared in state $\ket{1}$.}
\label{fig:pase_sf_u}
\end{center}
\end{figure*}

The implementation of the  terms written as products of the  $X$ and $P$ operators,
such as $H_{xp}=X^{u}_n P^{v}_n+P^{v}_n X^{u}_n$, 
is more difficult. Such terms appear in the squeezing Hamiltonian~(\ref{eq:xp_bsham}).
The problem is due to the fact that neither $\ket{x}$ nor
$\ket{p}$ are eigenvectors of $H_{xp}$. A general approach
for calculating $e^{-i \t H_{xp}}$ is the following.
First, the eigenproblem of the Hamiltonian $H_{xp}=\tilde{X}^{u} \tilde{P}^{v}+\tilde{P}^{v} \tilde{X}^{u}$
acting on the $N_x=2^{n_x}$ size space is solved numerically on a classical computer,
\begin{eqnarray}
\label{eq:Hxpeigene}
H_{xp}\ket{\nu}&=&E_\nu\ket{\nu}, \\
\label{eq:Hxpeigenu}
\ket{\nu} &=& \sum_{x=0}^{N_x-1} U_{\nu x} \ket{x}. 
\end{eqnarray}
\noindent  Second, since $U$ in Eq.~(\ref{eq:Hxpeigenu}) is unitary,  a circuit for
it can be designed~\cite{barenco_pra_1995, vartianien_2005}. 
The number of gates required for $U$ scales
as $4^{n_x}$.
The circuit for the evolution of $H_{xp}$ is shown in Fig.~\ref{fig:pase_sf_u}.
It starts  with the transformation $U$ applied to the boson register, which changes
the position vectors to the $H_{xp}$'s eigenvectors, $\ket{x} \xrightarrow{U} \ket{\nu}$. 
The next step is a gate yielding a phase factor $\exp(-i\t E_\nu)$. 
Since $E_\nu$ is calculated numerically a direct way to 
implement this is using $2^{n_x}$ multi($n_x$)-controlled CNOT gates acting on an ancilla qubit~\cite{footnote_ancilla}.
Because $n_x$-controlled gates scale as $\O{(n_x^2)}$~\cite{barenco_pra_1995}, 
the total number of gates required for the phase shift 
operation scales as $n_x^2 2^{n_x}$. 
The circuit ends with the transformation $U^{\dagger}$
applied to the boson register,  $\ket{\nu} \xrightarrow{U^{\dagger}} \ket{x}$.

Note that the circuits for the self-interacting boson terms  containing
$XP$ products scale exponentially with $n_x$. All the other circuits
for the fermion-boson model scale polynomial with $n_x$.
Nevertheless, remember that a $n_x$ exponential scaling is a polynomial scaling with
the boson cutoff number (since $N_{ph} \propto N_x$).
Moreover, the overall scaling of the full algorithm with the system size $N$ is not changed ({\em i.e.}, remains polynomial),  
since $n_x$ is constant (does not depend on $N$).

\subsection{Resource scaling}
\label{ssec:scaling}

Fermion-boson interacting systems are
represented on a number of qubits which scales linearly with the system size. 
Aside from the fermion-reserved qubits which scale linearly with the system size,
the number of additional  qubits required to represent the boson space is $\O(N n_x)$, where $N$ is the system size
and $n_x$ is the number of qubits necessary to map the 
low-energy Hilbert space of a single harmonic oscillator.
Equation~(\ref{eq:error}) implies that $n_x$  scales as 
\begin{equation}
\label{eq:nxscaling}
n_x = \log N_x = \O\left(\log \left[ \ln(\epsilon^{-1})
+0.765 N_{ph}\left( \epsilon^{-1} \right)\right]\right)
\end{equation}
where $\epsilon$ is the precision and $N_{ph}$ is the maximum boson cutoff number.
The boson cutoff number $N_{ph}$  is dependent on the target precision $\epsilon$.
 From Eq.~(\ref{eq:nxscaling}) one can infer that, as long as $N_{ph}$ increases
 with increasing $1/\epsilon$ slower than $\left[\ln(\epsilon^{-1})\right]^u$, with $u>0$
 being an arbitrary constant, then
\begin{equation}
\label{eq:nxscaling1}
n_x=\O(\log \left( \ln(\epsilon^{-1})\right).
\end{equation}
\noindent That is the case for problems discussed in Section~\ref{ssec:nph} where the boson states evolve under the action of an effective forced harmonic oscillator Hamiltonian; see Eq.~(\ref{eq:scnphe}).
Eqs.~(\ref{eq:scnphnzeta}), (\ref{eq:scnphne}) and (\ref{eq:nxscaling})
also imply that  $n_x$ scales as
\begin{eqnarray}
n_x&=&\O\left(\log \left(|\zeta_{max}|^2\right)\right),\\
n_x&=&\O\left(\log \left(N_E\right)\right),
\end{eqnarray}
where, as discussed in Section~\ref{ssec:nph}, $|\zeta_{max}|^2$ and $N_E$
are the effective coupling strength and the size of the low energy space under consideration, respectively.
For many problems, such as electron-phonon  models, we found that 
a  small number of qubits, $n_x\approx 6 \sim 7$,
is enough to accurately ($\epsilon<10^{-4}$) accommodate even the strong coupling regime.

Fermion-boson interacting systems can be simulated  in polynomial time. 
The estimation of the number of gates in the following analysis is for one Trotter step.

The number of gates and the circuit depth for simulating a single harmonic oscillator 
is $\O(n_x^2)$.
An $m$-body type boson-boson interaction ($m$-leg vertex interaction) term 
requires  $\O(n_x^m)$ gates. 
Boson self-interaction terms of type $XP$ (see Section~\ref{ssec:ev_bosonboson}) 
require $\O(4^{n_x})$ gates.

When the fermion-boson interaction and the boson-boson coupling have finite range, as is the case
for many physical models of interest in condensed matter physics,
the bosons  introduce an $\O(N)$ contribution to the total number of gates and a constant
contribution to the circuit depth.
For general long-range fermion-boson  interaction  the number of gates and circuit depth scale as $\O(N^2)$. 
When the bosons couple to the fermion hopping, as it is the case for electron-phonon models,
the  additional depth scales as  $\O(N)$.
For long-range $m$-leg vertex boson-boson interactions  the number of gates and circuit depth scale as $\O(N^m)$.

\subsection{Input state preparation}
\label{ssec:input_state}

The QPE algorithm requires an input state which has a large overlap with the
ground state of the system. The preparation of this state can be done using
the adiabatic method~\cite{Farhi_science_2001}. We start with a Hamiltonian 
$H_0=H_f+H_{h0}$, where $H_{h0}$ is the sum of the uncoupled harmonic oscillators,
and then slowly turn on the fermion-boson and boson-boson couplings. 
The ground state of $H_0$ is $\ket{f_0}  \otimes \ket{\Phi_0}$, where  
$\ket{f_0} $ is
the fermion Hamiltonian ground state. Its preparation, while non-trivial,  is addressed in 
the literature~\cite{ortiz_pra_2001,lidar_prl_2002, whitfield_2011,troyer_pra_2015}.
The state 
$\ket{\Phi_0}= \prod_n \otimes \ket{\tphi_{0n}}$ is the  ground state of  $H_{h0}$
and is a direct product of the harmonic oscillators ground state functions $\ket{\tphi_{0n}}$,
where $n$ is the  harmonic oscillator site label.
The state $\ket{\tphi_{0n}}$ is the zero$^{th}$-order HG function on the $2^{n_x}$ grid, {\em i.e.}, 
$\ket{\tphi_{0n}}=\ket{\chi_0}$, (see Eq.~(\ref{eq:dhg})). 
The preparation of $\ket{\Phi_0}$ therefore requires $N$ discrete Gaussian states in parallel, each state prepared on
a register of $n_x$ qubits.

Methods to prepare Gaussian states are discussed in Refs.~\cite{grover_wave,kitaev_gaussian}. 
These methods require quantum computation
of integrals, as well as arcosine and square root functions, with high precision, 
which might not be feasible 
on near-future computers with limited resources.

When $n_x$ is of the order of a few qubits, as it is for most electron-phonon problems of interest,
including our polaron example (Section~\ref{sec:polaron}, below), different methods can be employed 
to prepare a Gaussian on a grid. 

One choice is the brute force approach. 
The method is similar to the one described in Ref~\cite{grover_wave}, 
but the rotation angles for each configuration  are precomputed and implemented 
using $(n_x-1)$-controlled qubits. Since there are $2^{n_x}$ configurations and a
$(n_x-1)$-controlled qubits rotation scales as $\O(n_x^2)$~\cite{barenco_pra_1995}, the
corresponding circuit depth is $\O(n_x^2 2^{n_x})$. The circuit depth is independent of
the system size. 

Another possibility for small-register Gaussian state
preparation which is better suited for near-future quantum computers with limited coherence  time
is a variational method. We find heuristically that Gaussian states can be obtained with
high fidelity by applying a $N_S$-step unitary operator on the state $\ket{x=0}$,
\begin{eqnarray}
\label{eq:vinput}
\ket{\phi_v}= \prod_{s=1}^{N_S} U^s( \pmb{\theta}^s, \pmb{\rho}^s) \ket{x=0}, 
\end{eqnarray}
\noindent where 
\begin{eqnarray}
\label{eq:vprep_us}
&U^s( \pmb{\theta}^s, \pmb{\rho}^s)=\\ \nonumber
&=\prod_{i=0}^{n_x-1}\left(e^{-i \t_{yi}^s Y_i}  e^{-i \t_{xi}^s X_i} 
e^{-i \t_{zi}^s Z_i} \right) e^{-i \rho_{x}^s \X2}  e^{-i \rho_{p}^s \P2}.
\end{eqnarray}
\noindent The operators $e^{-i\rho^s_p \P2}$ and  $e^{-i\rho^s_x \X2}$ require
circuits with depth proportional to $n_x$, as described in Section~\ref{ssec:ev_ph}.
The single qubit  $x$, $y$ and $z$ rotations,   
$e^{-i \t_{x}^s X}$, $e^{-i \t_{y}^s Y}$ and $e^{-i \t_{z}^s Z}$ respectively, increase the circuit depth
by $3$ gates per step, since they can be implemented in parallel.
The varational parameters ${\pmb{\theta}}^s=\left\{\t^s_{xi}, \t^s_{yi}, \t^s_{zi} \right\}_{i=\overline{0,n_x-1}}$ and
${\pmb{\rho}}^s=\left\{\rho^s_{x}, \rho^s_{p}\right\}$ are determined by maximizing the fidelity
$|\bracket{\phi_v}{\chi_0}|^2$. 
The circuit depth for the  variational preparation of a Gaussian state is proportional 
to the number of steps $N_S$.

For large systems it is necessary to prepare  local Gaussian states with
high precision $\approx N^{-1}$. This can be understood from the following argument. 
If for a single harmonic oscillator the overlap between the prepared function and the Gaussian is   
$|\bracket{\phi}{\chi_0}|^2=1-\epsilon$, then the overlap of the wavefunction corresponding the $N$ harmonic oscillators  
with $\ket{\Phi_0}$
is $|\bracket{\phi}{\chi_0}|^{2N} \approx 1-N \epsilon$.

By employing  Eq.~\ref{eq:vinput} 
and the Simultaneous Perturbation Stochastic Approximation method~\cite{spall_1998}
for the optimization of $\left\{\pmb{\theta}^s, \pmb{\rho}^s\right\}$ parameters, 
we find that Gaussian states on $n_x=6, 7 \text{ and } 8$ qubit registers can be prepared
with fidelity larger than $0.988, 0.983 \text{ and } 0.981$,  respectively, in $N_S=3$  steps,
and fidelity larger than $0.999, 0.998 \text{ and } 0.996$, respectively, in 
$N_S=6$  steps. The fidelity values corresponding to $N_S=3$ ($N_S=6$)
are large enough for quantum computations of systems with $N \sim 100$ ($N \sim 1000$)
boson sites. 

A systematic investigation of the variational approach to Gaussian states preparation, 
including an analysis  of fidelity dependence on $n_x$ and $N_S$ 
remains to be addressed in a future study. Here we simply show that there is a practical
and efficient way to prepare  Gaussian states on small-register qubits.

\subsection{Measurements}
\label{ssec:meas}

Measurement methods for quantum algorithms simulating  many-particle systems, 
including local and time-dependent correlation functions have been described previously; see for example 
Refs.~\cite{somma_gubernatis_2002,troyer_pra_2015}. These methods can be applied to our
fermion-boson model as well. 

For example, to calculate the phonon distribution number in the Holstein polaron (in Section~\ref{sec:polaron})
we use QPE  for the unitary operator obtained by exponentiating
the boson number.

\section{Benchmarking the Holstein polaron on a quantum simulator}
\label{sec:polaron}

\begin{figure}
\begin{center}
\includegraphics*[width=3.4in]{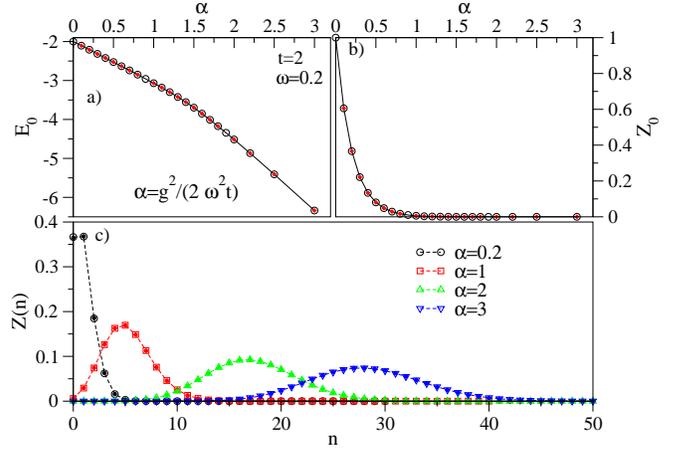}
\caption{The energy (a) and quasiparticle weight (b) for the $2$-site Holstein polaron versus electron-phonon coupling 
strength. 
(c) The phonon number distribution in the polaron state for different values of the coupling strength. 
The open symbols are computed
using the exact diagonalization technique while the full  small circles  are computed with using the QPE algorithm on a quantum simulator.
The state of each harmonic oscillator was stored in a   $n_x=6$ qubit register. }
\label{fig:2holstein}
\end{center}
\end{figure}

The bound state of an electron and its surrounding  phonons is called a polaron.
The polaron problem has been addressed extensively in the
literature. In the Holstein model~\cite{holstein_1959} the  phonons are described as a set
of independent oscillators located at every site. The  electron density at a particular site couples 
to the displacement of the harmonic oscillator located at the same site,
\begin{eqnarray}
\label{eq:ev_holstein}
H&=& -t \sum_{ij}  \left( c^{\dagger}_i c_j +c^{\dagger}_j c_i  \right) + g \sum_i c^{\dagger}_i c_i X_i\\ \nonumber
&+&\sum_i \frac{P_i^2}{2} +\frac{1}{2} \omega^2 X_i^2.
\end{eqnarray}

The $2$-site polaron problem is small enough to be solved exactly
using the diagonalization method on a conventional computer. The size of the local phonon Hilbert space
is truncated by a  cutoff large enough to reach convergence, {\em i.e.}, results no longer change with increasing cutoff. 
In our case a cutoff of $N_{ph} \approx 45$ phonons per site is sufficient for convergence.

In order to check the validity of our algorithm we ran a QPE code for the Holstein
polaron  on a $2$-site lattice using an Atos Quantum Learning Machine simulator. 
A comparison between
exact diagonalization and our quantum algorithm  is shown in Fig.~\ref{fig:2holstein}. 
The figure shows the energy of the polaron  as a function of the
dimensionless coupling constant $\alpha=g^2/2\omega^2t$, defined as a ratio of lattice deformation energy
$g^2/2 \omega^2$ to electron kinetic
energy $t$.
The agreement is very good, with a difference of $\O(10^{-4})$ due mainly to the use of the
Trotter approximation.
A register of size $n_x=6$ qubits for each harmonic oscillator is large enough to accurately describe the physics even 
in the strong coupling regime.

One can see that even this simple 2-site model
captures some essential features of more realistic polarons.
The transition from a light polaron to a heavy polaron as a function of the 
coupling strength is smooth, similar to what is 
seen in 1D polaron models~\cite{wellein_fehske}. 

The polaron state can be written as 
\begin{eqnarray}
\label{eq:ev_zdef}
\ket{\Phi}= \sum_{n=0} \sum_r a_{nr} \ket{n,r} 
\end{eqnarray}
\noindent  where $\{ \ket{n,r}\}_r$ are normalized vectors spanning the sector of the Hilbert space  with one electron and $n$ phonons.
The phonon distribution in the polaron state is defined as $Z(n)=\sum_r |a_{nr}|^2$
and can be determined by applying the QPE algorithm
for the phonon evolution Hamiltonian $H_{p}=\sum_i \frac{P_i^2}{2} +\frac{1}{2} \omega^2 X_i^2$.
Since $\ket{\Phi}$ is not an eigenstate of $H_{p}$, the energy $E_n=\omega(n+\frac{1}{2})$ is
measured with the probability $Z(n)$.

The quasiparticle weight $Z(0)$ as a function
of the coupling strength is  shown in Fig.~\ref{fig:2holstein} (b). This quantity represents the amount of the free electron in the polaron state
and gives the quasiparticle  weight measured in the photoemission experiments.
In Fig.~\ref{fig:2holstein} (c), $Z(n)$ is shown for several values of the coupling strength
corresponding to weak, intermediate and strong coupling regimes. 
The exact diagonalization and the QPE results agree well with each other.

Note that by employing a rotation of the oscillator coordinates such as, $X'_1=(X_1+X_2)/\sqrt{2}$
and $X'_2=(X_1-X_2)/\sqrt{2}$, the
two site Holstein polaron problem can be reduced to a single harmonic oscillator interacting with
a two-level system. However, we have not used this transformation to reduce the degrees of freedom of the problem,  since our purpose is benchmarking the algorithm.

\section{Discussion}
\label{sec:discuss}

Our algorithm assumes that, within a controlled approximation, each harmonic oscillator space 
can be truncated to a finite size. 
Since at every Trotter step the  state of each harmonic oscillator is a linear combination of displaced low energy states
(see Section~\ref{ssec:nph})
one expects that for electron-phonon models
the number of phonons in a localized basis can be truncated with exponential accuracy
as long as the system is stable.

Nevertheless, there are many physical system 
characterized by boson states with large occupancy where the implementation of
our algorithm might be questionable. In many cases, a convenient choice of the
boson basis might solve the truncation problem.
For example, it is evident that trying to employ the algorithm 
for a boson system having  the  zero momentum state macroscopically occupied,
as in case of Bose-Einstein condensates,  does not work if the harmonic oscillator labels
represent  momentum states. 
However, the truncation of the boson space might be feasible in the localized  basis since 
the number of  bosons per site is small.
In Appendix~\ref{app:bck} we show that 
the local occupation distribution of a zero-momentum macroscopically occupied state
goes  quickly to zero with increasing boson number.

Phase-squeezed states, commonly encountered in quantum optics, 
are another example of states with large boson number distributions.
Nevertheless, the  scaling equation for the boson register size 
derived for electron-phonon problems, Eq.~(\ref{eq:nxscaling1}), is also valid 
in this case. This can be inferred by noticing that the photon number distribution goes to zero 
exponentially quickly with increasing photon number. As shown in Ref.~\cite{garry_knight_book},
the probability $P_n$ to have $n$ photons in a squeezed state is proportional
to $\frac{C^n}{n!}$, where $C$ is independent of $n$. Using the Stirling formula $n! \approx \sqrt{2 \pi n} (n/e)^n$, one can see that
once $n >e C$, $P_n$ goes to zero faster than  exponentially with increasing $n$. However, $C$ can be large when the squeezing parameter is large.  In that case 
the squeezed state contains a large number of photons and, consequently,
its representation requires a large qubit register.

The possibility of implementing a simple and efficient displacement operator 
is the main merit of the boson space representation in our algorithm.
While the truncation of the boson space can be done more naturally in the boson number basis,
we are not aware of the existence of any efficient algorithm for the evolution 
of the forced harmonic oscillator in this basis.

Our work points to further directions of exploration.
Though we focus on electron-phonon interactions, our algorithm 
may be viewed as a first step towards simulating other interesting systems with these ingredients, 
QED and QCD among them. For example, the Hamiltonian representation of lattice gauge theories~\cite{Kogut:1974ag} 
possesses some structural similarities to the electron-phonon system. 
Pure gauge lattice Hamiltonians have been mapped onto quantum simulation schemes 
in~\cite{Byrnes:2005qx} and \cite{Zohar:2012xf} 
using different discretization schemes.
 It is interesting to consider applying this truncation of the Hilbert space to
more fundamental quantum field theories in order to understand the relationship between the cutoff in the number of bosons and regularization approaches.

\section{Conclusions}
\label{sec:concl}

We introduce a quantum algorithm for nonrelativistic fermion-boson interacting systems
which  extends the existing quantum fermion algorithms to include bosons.
The algorithm can address boson-boson interactions as well.
The bosons are represented as a set of harmonic oscillators.  Each harmonic oscillator space
is reduced to  a finite-sized Hilbert space $\dH$. We define  operators $\tilde{X}$ and $\tilde{P}$
on $\dH$ and show that, in the low-energy subspace, the algebra generated by $\{\tilde{X}, \tilde{P} \}$
is, up to an exponentially small error,  isomorphic with the algebra generated by $\{X, P \}$. 
The exponentially good representation is a consequence of the properties of the Hermite-Gaussian 
functions.  Since they fall exponentially fast to zero at large argument the Nyquist-Shannon theorem
can be applied. The Nyquist-Shannon theorem ensures both the truncation  of the HG functions to a finite set
of points and  the fact that the discrete Fourier transform is the same as the continuous Fourier transform at 
the grid points. The equivalence between the discrete and continuous Fourier transforms  makes possible 
the definition of the operators $\tilde{X}$ and $\tilde{P}$
obeying the canonical commutation relation in the low-energy subspace.
The size of the low-energy subspace is determined by the maximum order of the HG functions
whose width is covered by the grid size and it is equivalent to a cutoff in boson number $N_{ph}$. The minimum size of the 
finite space $\dH$ required to accurately represent $N_{ph}$ bosons 
increases approximately linearly with $N_{ph}$.

Our algorithm maps all harmonic oscillator  spaces  $\dH$  on the qubit space and
simulates the evolution operator of the fermion-boson Hamiltonian. 
The algorithm utilizes Trotter decomposition in small evolution steps. 
We present  circuits for the implementation of 
small  evolution steps corresponding to 
free boson, boson-boson interaction and fermion-boson interaction
terms in the Hamiltonian.

The number of qubits necessary to store the bosons scales logarithmically with the maximum boson 
number cutoff. For electron-phonons systems we find that a small number of qubits, $n_x \approx 6,7$ per harmonic oscillator, is large enough 
for the simulation of weak, intermediate and strong coupling regimes of most 
problems of physical interest.
The number of additional qubits required to add the boson space to a fermion system is $\O(N)$ where $N$ is proportional
to the system size.
For finite-range fermion-boson and boson-boson  interactions the bosons 
introduce a $\O(N)$ contribution to the total number of gates and a constant
contribution to the circuit depth. 
For general long-range fermion-boson  interactions the number of gates and circuit depth scale as $\O(N^2)$. 
When the bosons couple to the fermion hopping, as is the case for electron-phonon models,
the  additional depth scales as  $\O(N)$.
For long range $m$-leg vertex boson-boson interactions  the number 
of gates and circuit depth scale as $\O(N^m)$.

We benchmarked our algorithm on an Atos QLM  simulator for a two-site Holstein  polaron,  employing 
the QPE method. The  polaron energy  and phonon distribution 
are in excellent agreement with the ones calculated by exact diagonalization.

\section{ Acknowledgments}
We thank Andy Li, Eric Stern, Patrick Fox and Kiel Howe for discussions.
This manuscript has been authored by Fermi Research Alliance,
LLC under Contract No. DE-AC02-07CH11359 with the U.S. Department of Energy, Office of Science, Office of High Energy Physics. 
We gratefully acknowledge the computing resources provided and operated by 
the Joint Laboratory for System Evaluation (JLSE) at Argonne National Laboratory.
We would like to thank Atos for the use of their 38-Qubit Quantum Learning Machine (QLM) 
and support of their universal programming language AQASM.

\appendix
\section{Hermite-Gauss functions on a discrete grid }
\label{app:hor}

In this appendix we will show that, after the truncation to a discrete grid,
the discrete Fourier transform preserves 
the correspondence between the  direct and the Fourier transformed space of 
the  Hermite-Gauss functions with exponential precision.
This result is a consequence of the  Nyquist-Shannon sampling
theorem~\cite{nyquist-shanon} and the exponential falloff of the Hermite-Gauss (HG) functions at large argument.

The HG functions $\phi_n(x)$ (\ref{eq:HGx}) and their Fourier transforms $\hphi_n(p)$ (\ref{eq:fthg}) 
fall exponentially fast to zero for large argument. The width of the HG functions increases monotonically
with increasing $n$. Therefore for any positive integer cutoff $N_{ph}$, 
a half-width $L$ can be chosen such that  for all $n<N_{ph}$,
$|\hphi_n(p)| < \epsilon$ for  $|p|>L$ and  $|\phi_n(x)| < \epsilon$ for  $|x|>L$ ,
where $\epsilon \propto e^{-\frac{L^2}{2}}$ is exponentially small.

Let us define a periodic function $\overline{\phi}_n(p)=\phi_n(p)$ for $p \in [-L, L ]$ with the property
$\overline{\phi}_n(p)=\overline{\phi}_n(p+2L)$ and choose $L$ large enough such that 
\begin{eqnarray}
\label{eq:phiR}
\phi_n(p) = \overline{\phi}_n(p) R(\frac{p}{2L}) + \O(\epsilon),
\end{eqnarray}
\noindent where $R(t)$ is the rectangular function defined as
\begin{eqnarray}
\label{eq:R}
R(t)= \left\{ 
\begin{array}{ll}
1, &~ |t|\le \frac{1}{2}\\
0, &~ |t|>\frac{1}{2}
\end{array}
\right. .
\end{eqnarray}
\noindent Since the function $\phi_n(x)$ is the Fourier transform $\phi_n(p)$, according to Eq.~(\ref{eq:phiR})
it can be written as a convolution of  $\overline{\phi}_n(x)$ and $v(x)$, 
\begin{eqnarray}
\label{eq:phixcv}
\phi_n(x) &=& \int \overline{\phi}_n(y) v(x-y) dy + \O(\epsilon) \\ \nonumber
&=& \sum^{\infty}_{i=-\infty} \overline{\phi}_n(x_i) v(x-x_i)+ \O(\epsilon).
\end{eqnarray}
\noindent In Eq.~(\ref{eq:phixcv})
\begin{eqnarray}
\label{eq:phipft}
    \overline{\phi}_n(x_i)&=& \frac{1}{2L}\int^L_{-L} \overline{\phi}_n(p) e^{ip x_i} dp
    \\ \nonumber
    &= &\frac{1}{2L}\phi_n(x_i)+ \O(\epsilon),
\end{eqnarray}
\noindent is the Fourier transform of the periodic function $\overline{\phi}_n(p)$ 
defined for the discrete set of points $\{x_i= i \Delta \}_{i\in \mathbb{Z}}$,
where $\Delta=\frac{\pi}{L}$.
The function $v(x)$ is the Fourier transform of the rectangular function $R(\frac{p}{2L})$, 
\begin{eqnarray}
\label{eq:uft}
    v(x)&=&\int R(\frac{p}{2L}) e^{ip x} dp = \int^L_{-L} e^{ip x} dp
    \\ \nonumber
    &=& 2\frac{\sin Lx}{x}.
\end{eqnarray}
From Eqs.~(\ref{eq:phixcv}),  (\ref{eq:phipft}) and (\ref{eq:uft}) 
one can write
\begin{eqnarray}
\label{eq:dhg_sampling}
\phi_n(x)=\sum_{i=-\infty}^{\infty} \phi_n(x_i) u_i(x)+ \O(\epsilon),
\end{eqnarray}
\noindent with $u_i(x)= \frac{1}{2L}v(x-x_i)$.

In fact, if the exponentially small term  $\O(\epsilon)$ is neglected, 
Eq.~(\ref{eq:dhg_sampling}) is the well-known  Nyquist-Shannon sampling
theorem~\cite{nyquist-shanon} which states that a  band-limited function 
can be sampled without loss of information 
at points $x_i=i \frac{\pi}{L}$, where $L$ is the band half-width.
Note that $u_i(x)=\sinc\left(\frac{x-x_i}{\Delta}  \right)$ where 
$\sinc(x)=\frac{\sin (\pi x) }{\pi x}$, is the normalized sinc function, which is familiar in
signal processing.

Since $\phi_n(x)$ is also exponentially small for $|x|>L$, the summation over $i$
in Eq.~(\ref{eq:dhg_sampling}) can be truncated to the interval $[-L, L]$.  
The minimum number of points $N_x$ necessary to 
sample the interval $[-L, L]$ should satisfy the equation
$N_x \Delta =2 L$.  Since $\Delta=\frac{\pi}{L}$,
it follows  that $2L=\sqrt{2 \pi N_x}$  and $\Delta=\sqrt{\frac{2 \pi}{N_x}}$.

The functions $u_i(x)$ appearing in Eq.~(\ref{eq:dhg_sampling}) form an orthonormal
set,
\begin{eqnarray}
\label{eq:uiuj}
\int u_i(x) u_j(x) dx =  2 \pi \int \hat{u}^*_i(p) \hat{u}_j(p) dp=  \Delta \delta_{ij},
\end{eqnarray}
\noindent as can be easily checked by noticing that
 \begin{eqnarray}
\label{eq:sinc}
\hat{u}_i(p)=\frac{1}{2 \pi} \int  u_i(x) e^{-ipx} dx= \frac{1}{2L} e^{-ipx_i} R (\frac{p}{2 L}).
\end{eqnarray}
Eqs.~(\ref{eq:dhg_sampling}) and (\ref{eq:uiuj}), together with the orthogonality property of the HG functions, imply
 \begin{eqnarray}
\label{eq:dhg_orto}
     \delta_{nm}&=&\int \phi_n(x) \phi_m(x) dx
     \\ \nonumber 
     &=&\Delta \sum_{i=-\frac{N_x}{2}}^{\frac{N_x}{2}-1} \phi_n(x_i)  \phi_m(x_i),
\end{eqnarray}
\noindent where we neglect writing the exponentially small term $\O(\epsilon)$. Consequently, the $N_x$ size vectors 
\begin{eqnarray}
\label{eq:adhg}
\chi_n(x_i)= \sqrt{\Delta} \phi_n(x_i),
\end{eqnarray}
\noindent form an orthonormal set. Note that this is not a complete  set since  Eq.~(\ref{eq:dhg_orto}) is
valid only for  $n,m<N_{ph}<N_x$. The properties of the HG functions require $N_{ph}<N_x$,
as implied by Eq.~(\ref{eq:NxNph}).

The Fourier transform of the HG functions can be written as
\begin{eqnarray}
\label{eq:hg_ft}
\hat{\phi}_n(p)&=&\sum_{i=-\frac{N_x}{2}}^{\frac{N_x}{2}-1} \phi_n(x_i) \hat{u}_i(p) \\ \nonumber
&=&\frac{1}{2L} \sum_{i=-\frac{N_x}{2}}^{\frac{N_x}{2}-1}\phi_n(x_i) e^{-i p x_i} R(\frac{p}{2L}).
\end{eqnarray}
\noindent Since $R(\frac{p}{2L})=1$ for $p \in [-L,L]$, up to an exponentially small error one can write
\begin{eqnarray}
\label{eq:adfthg}
\hat{\phi}_n(p_m)= \frac{1}{2L} \sum_{i=-\frac{N_x}{2}}^{\frac{N_x}{2}-1} \phi_n(x_i)e^{-i p_m x_i},
\end{eqnarray}
\noindent where $p_m=m \Delta$ with $m=\overline{-N_x/2, N_x/2-1}$. This implies
\begin{eqnarray}
\label{eq:adft}
\hat{\chi}_n(p_m)=\sqrt{2 \pi \Delta} \hat{\phi}_n(p_m),
\end{eqnarray}
\noindent where
\begin{eqnarray}
\label{eq:adft1}
\hat{\chi}_n(p_m)= \frac{1}{\sqrt{N_x}} \sum_{i=-\frac{N_x}{2}}^{\frac{N_x}{2}-1} \chi_n(x_i)e^{-i p_m x_i},
\end{eqnarray}
\noindent is the discrete Fourier transform of the finite size vector $\chi_n(x_i)$ defined in Eq.~(\ref{eq:adhg}).
Equation (\ref{eq:adft}) shows that, 
when restricted to the grid points $\{x_i \} \longleftrightarrow \{p_m \}$,
the discrete Fourier transform can replace the continuous
Fourier transform of the HG functions with $n<N_{ph}$ with exponentially small error.

\section{Coherent states, displacement operator and the forced harmonic oscillator}
\label{app:fho}

We list here some properties of boson coherent states and the forced harmonic oscillator
relevant for the  boson number cutoff discussion presented in Section~\ref{ssec:nph}. 
The derivation of these results can be found in Refs.~\cite{garry_knight_book,cahill_pr_1969, oliveira_pra_1990, merzbacher}.

The displacement operator is defined as   
\begin{equation}
\label{eq:dis}
D(z)=e^{z b^{\dagger}-z^* b },
\end{equation}
\noindent where $z$ is a complex number. The displacement operator
has the following properties

\begin{eqnarray}
\label{eq:dzbdz}
D^{\dagger}(z)b D(z) &=& b+z \\
\label{eq:dzbddz}
D^{\dagger}(z)b^{\dagger} D(z)&=&b^{\dagger}+z^*\\
\label{eq:dz1z2}
D(z_1) D(z_2) &=& e^{\frac{z_1 z^*_2-z^*_1 z_2}{2} } D(z_1+z_2).
\end{eqnarray}

When applied to the vacuum, $D(z)$ creates  the coherent state
\begin{equation}
\label{eq:chs}
\ket{z} \equiv D(z)\ket{0}= e^{-\frac{\left|z\right|^2}{2}} \sum_{n=0}^\infty \frac{z^n}{\sqrt{n!}}\ket{\phi_n}.
\end{equation}

The coherent states are eigenstates of the annihilation operator
and have the following properties
\begin{eqnarray}
\label{eq:bz}
b\ket{z}&=&z\ket{z}\\
\label{eq:=omegaz}
e^{-i \t b^{\dagger} b} \ket{z}&=&\ket{ze^{-i \t}}\\
\label{eq:n}
\langle N \rangle &=&\langle b^{\dagger} b \rangle =\left|z\right|^2\\
\label{eq:sigman}
\langle (\Delta N)^2 \rangle &=&\langle \left(N-\langle N \rangle \right)^2 \rangle =\left|z\right|^2 .
\end{eqnarray}
\noindent The  boson occupation number  of a coherent state
is a Poisson distribution (see Eq.~(\ref{eq:chs})), 
\begin{eqnarray}
\label{eq:zn}
\left| \bracket{\phi_n}{z} \right|^2 &=& e^{-\left|z\right|^2}\frac{\left|z\right|^{2n}}{n!}.
\end{eqnarray}
\noindent This distribution falls exponentially to zero with increasing $n$. The coherent states
can be represented with accuracy $\epsilon$ on a finite-sized space 
truncated by a boson number cutoff 
$N_{c}\gtrsim |z|^2+|z| \sqrt{2 \ln(\epsilon^{-1})}$.

The displacement  matrix in the boson number basis is
\begin{equation}
\label{eq:mdn}
\bra{\phi_m} D(z)\ket{\phi_n}=\sqrt{\frac{n!}{m!}}z^{m-n}e^{-\left|z\right|^2/2} L_n^{(m-n)}(\left|z\right|^2),
\end{equation}
\noindent where $L_n^{(m-n)}$ are the Laguerre polynomials~\cite{cahill_pr_1969,oliveira_pra_1990}.
The {\em displaced number states} are defined by applying the displacement operator to the 
number states,
\begin{eqnarray}
\label{eq:dzn}
\ket{n,z}=D(z)\ket{\phi_n}.
\end{eqnarray}
A useful property of the displaced number states is
\begin{eqnarray}
\label{eq:H0dzn}
e^{-i\t b^{\dagger} b}\ket{n,z}= e^{-i n\t } \ket{n, z e^{-i\t}},
\end{eqnarray}
\noindent which can be obtained by expanding $\ket{n,z}$ in the $\{\ket{\phi_m}\}_m$ basis and employing Eq.~(\ref{eq:mdn}).

Using Eq.~(\ref{eq:mdn}) one can show that the boson  number distribution
in the displaced number state  $\ket{n,z}$ is Poissonian (see Ref.~\cite{oliveira_pra_1990} for more details) 
with the mean and variance given by 
\begin{eqnarray}
\label{eq:mean_nz}
\langle N \rangle &=&n+\left|z\right|^2 \\
\label{eq:var_nz}
\langle (\Delta N)^2 \rangle &=& \left(2n+1\right)\left|z\right|^2.
\end{eqnarray}
\noindent  The  state $\ket{n,z}$ can be represented to accuracy $\epsilon$ on a finite-sized space 
truncated by a boson number cutoff 
\begin{equation}
\label{eq:nz_cut}
N_{c}\gtrsim n+|z|^2+|z| \sqrt{2 (2 n+1) \ln(\epsilon^{-1})}.
\end{equation}

The forced harmonic oscillator Hamiltonian
\begin{eqnarray}
\label{eq:fhoxp_ham}
H= \frac{P^2}{2M} + \frac{1}{2}M \omega^2+g(t) X +h(t) P
\end{eqnarray}
\noindent is written in the second quantized form as
\begin{eqnarray}
\label{eq:fho_ham}
H=  \omega \left( b^{\dagger} b +\frac{1}{2} \right) + f(t) b+ f^*(t) b^{\dagger},
\end{eqnarray}
\noindent with $f(t)= [g(t)/\sqrt{M \omega}-i h(t) \sqrt{M \omega}]/\sqrt{2} $.

When $f$ is constant, the forced harmonic oscillator is, up to a constant term $|f|^2/\omega$,
the harmonic oscillator $H_0=\omega  ( b^{\dagger} b +1/2 )$ displaced by $f^*/\omega$, 
\begin{eqnarray}
D^{\dagger}  \left(\frac{f^*}{\omega}\right) H_0 D \left(\frac{f^*}{\omega}\right)=H+\frac{\left| f\right|^2}{\omega}.
\end{eqnarray}
\noindent Therefore the eigenstates of  $H$ can be obtained from the eigenstates of $H_0$
by applying $D(f/\omega)$, {\em i.e.}, are displaced number states, $\{\ket{n,f/\omega}\}_n$,
as defined by Eq.~(\ref{eq:dzn}).

One can show that the evolution operator of the Hamiltonian (\ref{eq:fho_ham}) is~\cite{merzbacher}
\begin{eqnarray}
\label{eq:fho_ev}
U(t)=e^{i \beta(t)} D\left[\zeta(t) e^{ -i\omega t}\right] e^{-i H_0 t}
\end{eqnarray}
\noindent with
\begin{eqnarray}
\label{eq:beta_ev}
\beta(t)&=&\frac{i}{2}\int_0^t du \int_0^u ds f(u) f^*(s) e^{- i\omega \left(u-s \right)} \\ \nonumber
&&-\frac{i}{2}\int_0^t du \int_0^u ds f^*(u) f(s) e^{ i\omega \left(u-s \right)}, \\
\end{eqnarray}
\noindent and
\begin{eqnarray}
\label{eq:zeta_ev}
\zeta(t)&=& -i \int^t_0 f^*(u) e^{i \omega u} du.
\end{eqnarray}
Employing Eqs.~(\ref{eq:dz1z2}), (\ref{eq:H0dzn}) and (\ref{eq:fho_ev}), the evolution of a displaced number state is
\begin{equation}
\label{eq:app_ugnz}
U(t)\ket{n,z}= e^{i \gamma} e^{i\left(\beta-n \omega t\right)} \ket{n, \left(\zeta +z\right)e^{-i \omega t}},
\end{equation}
where $\gamma= -i(\zeta z^*-\zeta^* z)/2$ is a real phase factor.

A  state belonging to the finite-sized space determined by the $N_E$ cutoff
\begin{eqnarray}
\label{eq:phi_tr}
\ket{\phi}=\sum^{N_E}_{n=0} c_n \ket{\phi_n},
\end{eqnarray}
\noindent evolves under the forced harmonic oscillator action as
\begin{eqnarray}
\label{eq:uphi}
U(t)\ket{\phi}&=& \sum^{N_E}_{n=0} c_n e^{i\left(\beta-n \omega t\right)} \ket{n, \zeta e^{-i \omega t}}.
\end{eqnarray}
\noindent Provided that $|\zeta(t)|_t<\zeta_{max}$, the  state  $\ket{\phi}$ remains contained in the low-energy
subspace defined by the new cutoff 
\begin{equation}
\label{eq:nphscaling}
N_{ph}=N_E+|\zeta_{max}|^2 +|\zeta_{max}| \sqrt{2 (2 N_E+1) \ln(\epsilon^{-1})},
\end{equation}
with $\epsilon$ accuracy.

\section{Local boson distribution of a macroscopically occupied state with zero momentum}
\label{app:bck}

In this Appendix we demonstrate 
that a zero-momentum state occupied by $N$ bosons, where $N$ is the system size, can be truncated to a finite number of bosons
with exponential accuracy  when $N \longrightarrow \infty$ 
if the truncation of the Hilbert space is done in the localized basis. 

Let us consider $N$ bosons on a lattice of size $N$. The boson creation operators in the momentum basis $\{\tilde{b}_k^{\dagger}\}$ 
are related to the boson creation operators in the localized basis $\{b_n^{\dagger}\}$ by
\begin{eqnarray}
\label{eq:bk}
\tilde{b}_k^{\dagger}=\frac{1}{\sqrt{N}}\sum^{N-1}_{n=0} e^{i k r_n} b_n^{\dagger}.
\end{eqnarray}
\noindent  Next, we consider the  state  where all $N$ particles have zero momentum,
\begin{eqnarray}
\label{eq:bes}
\ket{\phi}\equiv \ket{N_{k=0}}= \frac{1}{\sqrt{N!}} \left(\tilde{b}_0^{\dagger}\right)^N \ket{0}.
\end{eqnarray}

Now we ask  what is the probability of having $p$ bosons occupying a localized state labeled  $i$. 
Expanding in a localized basis, the  state $\ket{\phi}$ can be written as
\begin{eqnarray}
\label{eq:bes1}
\ket{\phi} = \sum_{p=0}^N w(p) \ket{p_i} \ket{\left(N-p\right)_a},
\end{eqnarray}
\noindent where 
\begin{eqnarray}
\label{eq:pi}
\ket{p_i}= \frac{1}{\sqrt{p!}} \left(b_i^{\dagger}\right)^p \ket{0},
\end{eqnarray}
\noindent is the state with $p$ bosons at site $i$, and  $\ket{\left(N-p\right)_a}$ is a site with $N-p$ bosons
anywhere else. 

The calculation of the probability $|w(p)|^2$ for having $p$ bosons on the site $i$ is straightforward.
If one defines the boson operator 
\begin{eqnarray}
\label{eq:ai}
a^{\dagger}=\frac{1}{\sqrt{N-1}}\sum^{N-1}_{\substack{n=0 \\ n \ne i  } } b_n^{\dagger}, 
\end{eqnarray}
\noindent then the $k=0$ creation operator can be written as
\begin{eqnarray}
\label{eq:bkzero}
\tilde{b}_0^{\dagger}=\frac{1}{\sqrt{N}}\sum^{N-1}_{n=0} b_n^{\dagger} =\frac{1}{\sqrt{N}}\left(b_i^{\dagger} + \sqrt{N-1}a^{\dagger} \right).
\end{eqnarray}
\noindent Introducing  Eq.~(\ref{eq:bkzero}) into Eq.~(\ref{eq:bes}) one obtains
\begin{eqnarray}
\label{eq:bes2}
\ket{\phi}&=& \frac{1}{\sqrt{N!}} \left( \frac{1}{\sqrt{N}} \right)^N \sum_{p=0}^{N} \\ \nonumber
&&\binom{N}{p} \left(\sqrt{N-1}\right)^{N-p}\left( b_i^{\dagger} \right)^p \left( a^{\dagger} \right)^{N-p} \ket{0} \\ \nonumber
&=& \left( \frac{\sqrt{N-1}}{\sqrt{N}}\right)^N \sum_{p=0}^{N} \sqrt{ \binom{N}{p}} \frac{1}{\sqrt{N-1}^p} \ket{p_i}\ket{\left(N-p\right)_a},
\end{eqnarray}
\noindent where 
\begin{eqnarray}
\label{eq:Npa}
\ket{\left(N-p\right)_a}= \frac{1}{\sqrt{\left(N-p\right)!}} \left(a^{\dagger}\right)^{N-p}\ket{0}.
\end{eqnarray}

It follows that
\begin{eqnarray}
\label{eq:pphi}
|w(p)|^2 &=& \left(\frac{N-1}{N}\right)^N \frac{N (N-1)...(N-p+1)}{p! \left(N-1\right)^p}  \\ \nonumber
&<& \frac{1}{p!} \left(\frac{N-1}{N} \right)^{N-1} \xrightarrow{N \longrightarrow \infty} \frac{1}{p! e}.
\end{eqnarray}
\noindent The occupation number probability at a particular site falls as $1/p!$  with increasing the occupation number $p$.
Therefore, the state $\ket{\phi}$  can be truncated with exponential accuracy 
on the Hilbert space  obtained as a product of finite-sized local Hilbert spaces.


\begin{thebibliography}{99}
\bibitem{footnote_ancilla} Likely, an optimal implementation of the phase factor $\exp(-i\t E_\nu)$
can be achieved without using ancilla qubits.
\end{thebibliography}
\end{document}